\documentclass[fleqn,usenatbib]{mnras}

\usepackage{newtxtext,newtxmath}

\usepackage[T1]{fontenc}

\DeclareRobustCommand{\VAN}[3]{#2}
\let\VANthebibliography\thebibliography
\def\thebibliography{\DeclareRobustCommand{\VAN}[3]{##3}\VANthebibliography}

\usepackage{graphicx}	
\usepackage{subcaption}	%
\usepackage{amsmath}	
\usepackage{siunitx}    
\usepackage{multirow}

\newcommand{\guvb}{\Gamma_{\text{HI}}}
\newcommand{\lya}{Ly$\alpha$\ }
\newcommand{\lammfp}{\lambda_{\text{mfp}}}
\newcommand{\snr}{\text{SNR}_{10}}

\DeclareSIUnit \h {h}
\DeclareSIUnit \parsec {pc}

\title[Forecasting $\lammfp$ constraints]{Forecasting constraints on the mean free path of ionizing photons at $z \geq 5.4$ from the Lyman-$\alpha$ forest flux auto-correlation function
}

\author[Molly Wolfson et al.]{
Molly Wolfson$^{1}$\thanks{E-mail: mawolfson@ucsb.edu},
Joseph F. Hennawi$^{1,2}$,
Frederick B. Davies$^{3}$,
and Jose O{\~{n}}orbe$^{4}$
\\
$^{1}$Department of Physics, University of California, Santa Barbara, CA 93106, USA\\
$^{2}$Leiden Observatory, Leiden University, Niels Bohrweg 2, 2333 CA Leiden, Netherlands\\
$^{3}$Max-Planck-Institut f\"{u}r Astronomie, K\"{o}nigstuhl 17, 69117 Heidelberg, Germany\\
$^{4}$Facultad de F\'{i}sicas, Universidad de Sevilla, Avda. Reina Mercedes s/n, Campus de Reina Mercedes, 41012 Sevilla, Spain
}

\date{Accepted XXX. Received YYY; in original form ZZZ}

\pubyear{2022}

\begin{document}
\label{firstpage}
\pagerange{\pageref{firstpage}--\pageref{lastpage}}
\maketitle

\begin{abstract}
Fluctuations in Lyman-$\alpha$ (Ly$\alpha$) forest transmission towards high-$z$ quasars are partially sourced from spatial fluctuations in the ultraviolet background (UVB), the level of which are set by the mean free path of ionizing photons ($\lammfp$).
The auto-correlation function of Ly$\alpha$ forest flux characterizes the strength and scale of transmission fluctuations and, as we show, is thus sensitive to $\lammfp$.
Recent measurements at $z \sim 6$ suggest a rapid evolution of $\lammfp$ at $z>5.0$ which would leave a signature in the evolution of the auto-correlation function.
For this forecast, we model mock Ly$\alpha$ forest data with properties similar to the XQR-30 extended data set at $5.4 \leq z \leq 6.0$.
At each $z$ we investigate 100 mock data sets and an ideal case where mock data matches model values of the auto-correlation function. 
For ideal data with $\lammfp=9.0$ cMpc at $z=6.0$, we recover $\lammfp=12^{+6}_{-3}$ cMpc.
This precision is comparable to direct measurements of $\lammfp$ from the stacking of quasar spectra beyond the Lyman limit.
Hypothetical high-resolution data leads to a $\sim40\%$ reduction in the error bars over all $z$.
The distribution of mock values of the auto-correlation function in this work is highly non-Gaussian for high-$z$, which should caution work with other statistics of the high-$z$ Ly$\alpha$ forest against making this assumption. 
We use a rigorous statistical method to pass an inference test, however future work on non-Gaussian methods will enable higher precision measurements. 
\end{abstract}

\begin{keywords}
intergalactic medium -- dark ages, reionization, first stars -- quasars: absorption lines -- methods: statistical
\end{keywords}



\section{Introduction}

The neutral hydrogen in the intergalactic medium (IGM) was reionized by the first luminous sources during the epoch of reionization. 
This period was one of the most dramatic changes in the history of the universe. 
Current Planck constraints from the cosmic microwave background put the midpoint of reionization at $z_{\text{re}} = 7.7 \pm 0.7$ \citep{planck_2018}. 
There have also been multiple measurements that suggest reionization was not completed until after $z \leq 6$ \citep{fan_2006, becker_2015, becker_2018, bosman_2018, bosman_2021_data, eilers_2018, boera_2019, yang_2020, jung_2020, kashino_2020, moreales_2021}.
However, much is still unknown about this process such as the exact timing, the impact on the thermal state of the IGM, the driving sources, and the number of photons that must be produced to complete reionization. 

Characterizing the IGM both during and immediately after reionization will give vital information to answer these remaining questions. 
Of particular interest is the average distance that the ionizing photons travel through the IGM before interacting with its neutral hydrogen -- also known as the mean free path of ionizing photons, $\lammfp$. 
The end of reionization results in a rapid increase in $\lammfp$ as the initially isolated regions of ionized hydrogen overlap to form a mostly ionized universe \citep{gnedin_2000, gnedin_fan_2006, wyithe_2008, daloisio_2018, kulkarni_2019, keating_2020_a, keating_2020_b, nasir_daloisio_2020, cain_2021, gnedin_2022}. 
Detecting this rapid increase is therefore a clear signal of the end of reionization.

Direct measurements of $\lammfp$ at $z \leq 5.2$ have been achieved from stacked quasar spectra \citep{prochaska_2009, fumagalli_2013, omeara_2013, worseck_2014}. 
Using a similar method, \citet{becker_2021} recently reported measurements of $\lammfp = 9.09^{+1.62}_{-1.28}$ proper Mpc at $z = 5.1$ and $\lammfp = 0.75^{+0.65}_{-0.45}$ proper Mpc at $z = 6$. 
This value at $z=6$ is significantly smaller than extrapolations from previous lower $z$ measurements \citep{worseck_2014}, causes tension with measurements of the ionizing output from galaxies \citep{cain_2021, davies_2021}, and also suggests a roughly 12-fold increase in $\lammfp$ between $z = 6$ and $z = 5.1$, potentially signalling the end of reionization.
An alternative method presented in \citet{bosman_2021_limit} used lower limits on individual free paths towards high-$z$ sources to place a $2 \sigma$ limit of $\lammfp > 0.31$ proper Mpc at $z = 6.0$. 
This \citet{bosman_2021_limit} method is similar to other measurements using individual free paths \citep{songaila_2010, rudie_2013, romano_2019}.
Additional independent methods of measuring $\lammfp$ are necessary to verify these measurements. 
Of particular interest are methods that can be used at several redshift bins at $z > 5$ in order to study the evolution of $\lammfp$ in finer detail.

In this paper we investigate using the auto-correlation function of \lya forest flux in high-$z$ quasar sightlines to constrain $\lammfp$. 
The Ly$\alpha$ opacity, $\tau_{\rm Ly\alpha}$, is related to $\lammfp$ via $\tau_{\rm Ly\alpha} = n_{\rm HI} \sigma_{{\rm Ly}\alpha} \propto  1 / \Gamma_{\rm HI} \propto 1 / \lammfp^{\alpha}$ where $\alpha$ is typically between 3/2 and 2 (see e.g. \citet{Rauch_1998, haardt_madau_2012}). 
Additionally, during reionization the existence of significant neutral hydrogen in the IGM will cause a short mean free path value to also result in large spatial fluctuations in the  ultraviolet background (UVB). 
This is because, during reionization ionizing photons are produced from the first sources and then quickly absorbed by the remaining neutral hydrogen. 
Thus there are large values of the UVB where the photons are produced and very small values where neutral hydrogen remains. 
If the mean free path is large, photons will travel further and effectively smooth the UVB \citep{mesinger_furlanetto_2009}.
The positive fluctuations in the UVB on small scales that accompany a short mean free path would then boost the flux of the \lya forest on small scales, which could then be detected in the auto-correlation function.
Various previous studies have investigated the effect of large scale variations in the UVB on the auto-correlation function and power spectrum of the \lya forest \citep{zuo_1992_a, zuo_1992_b, croft_2004, meiksin_2004, mcdonald_2005, gontcho_2014, pontzen_2014_a, pontzen_2014_b, daloisio_2018, meiksin_2019, onorbe_2019}. 
Our work is focused on determining if the effect of the fluctuating UVB on the auto-correlation function can lead to a constraint on $\lammfp$.  

While the power spectrum has been a more popular statistic used on the high-$z$ \lya forest to date \citep{boera_2019, walther_2019, gaikwad_2021}, the auto-correlation function has a few characteristics that make it easier to work with than the power spectrum.
The two most obvious are the effect of noise and masking on the auto-correlation function when compared to the power spectrum. 
Astronomical spectrograph noise is expected to be white or uncorrelated. 
Uncorrelated noise only impacts the auto-correlation function at zero lag, since at all other lags the uncorrelated noise will average to zero. 
Therefore, by not measuring the auto-correlation at zero lag we have fully removed the effect of white noise. 
On the other hand, white noise is a constant positive value at all scales for the power spectrum. 
Thus the unknown noise level must be calculated and subtracted from power spectrum measurements which will add additional uncertainty to the final measurement. 
Additionally, real data often has regions of spectra that need to be removed from the quasar spectrum (e.g. for metal lines). 
Masking out these and other regions introduces a complicated window function to the power spectrum that must be corrected for (see e.g. \citet{walther_2019}) and will again increasing the uncertainty in the measurement. 
The auto-correlation function does not require a similar correction since masking only result in fewer points in bins for certain lags. 

The structure of this paper is as follows. 
We discuss our simulation data in Section \ref{section: sim data}. 
The auto-correlation function and our other statistical methods are described in Section \ref{section: methods}. 
We then discuss our results in Section \ref{section: results} and summarize in Section \ref{section: conclusion}. Here we also touch on how additional work on modeling $\lammfp$ in simulations as well as better statistical methods will improve these constraints.

\section{Simulation Data} \label{section: sim data}

\subsection{Models} \label{section: sim data uvb model}

In this work we use a simulation box run with \texttt{Nyx} code \citep{almgren_2013}. 
\texttt{Nyx} is a hydrodynamical simulation code that was designed for simulating the Ly$\alpha$ forest with updated physical rates from \citet{lukic_2015}.
The \texttt{Nyx} box has a size of $L_{\text{box}} = 100$ cMpc $h^{-1}$ with $4096^3$ dark matter particles and $4096^3$ baryon grid cells. 
This box is reionized by a \citet{haardt_madau_2012} uniform UVB that is switched on at $z \sim 15$. 
We have two snapshots of this simulation at $z = 5.5$ and $z = 6$. 
In this work we want to consider these models at seven redshifts: $5.4 \leq z \leq 6$ with $\Delta z = 0.1$.
In order to consider the redshifts for which we do not have a simulation output, we select the nearest snapshot and use the desired redshift when calculating the proper size of the box and the mean density. 
This means we use the density fluctuations, temperature, and velocities directly from the nearest \texttt{Nyx} simulation output. 
We additionally used the $z=6.0$ simulation snapshot to generate low-resolution skewers at $z = 5.7$ and found no significant change in our finally results, confirming that using the nearest simulation snapshot in this way is sufficient.

\begin{figure}
	\includegraphics[width=\columnwidth]{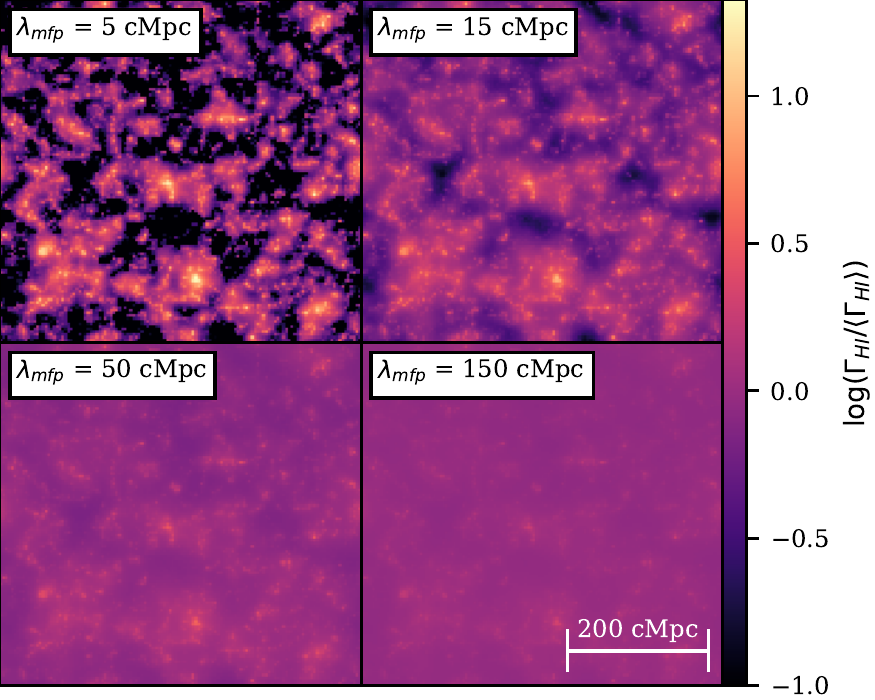}
    \caption{
        Each quadrant of this figure shows a slice through the box of the $z = 5.5$ UVB model used for four example values of $\lammfp$ (5, 15, 50, and 150 cMpc). 
        The colorbar is cut off at $\log(\guvb / 10^{-12} \SI{}{\per\second}) = -1$ in order to better visualize the differences between the models. 
        The models with smaller $\lammfp$ values show greater variation in the UVB than those with larger $\lammfp$, as expected. 
    }
    \label{fig:mfp_slices}
\end{figure}

\begin{figure}
	\includegraphics[width=\columnwidth]{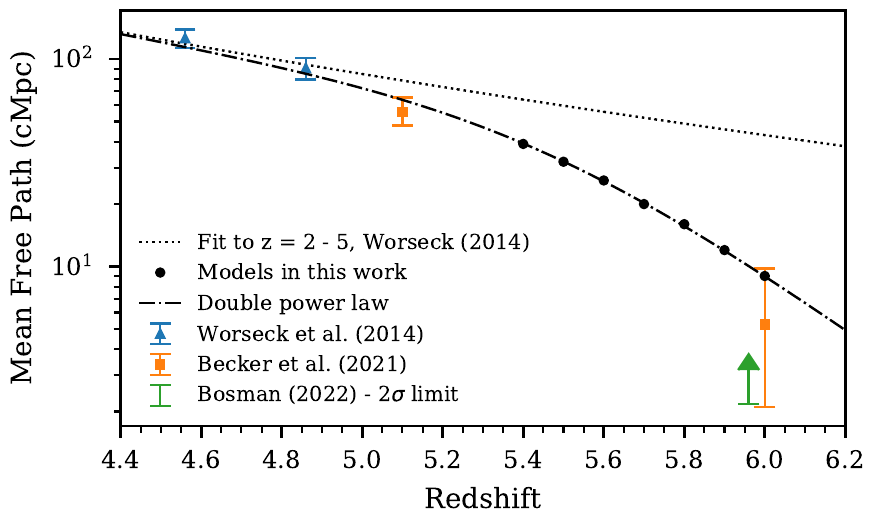}
    \caption{
        The blue triangles and orange squares show previous measurements of $\lammfp$ at high-$z$ from \citet{becker_2021} and \citet{worseck_2014} respectively. 
        The green limit is from \citet{bosman_2021_limit}. 
        Additionally, the dotted line shows the results of the power law fit to data from $z = 2 - 5$ from \citet{worseck_2014}. 
        For this work, we modified this power law fit into a double power law using the same low-$z$ scaling by eye in order to agree with the \citet{becker_2021} points. 
        This new scaling is shown by the dot-dashed line. 
        We used this double power law as an example redshift evolution of $\lammfp$, where the values we modeled are shown as black circles. 
    }
    \label{fig:mfp_model_evolution}
\end{figure}

We also have separate boxes of fluctuating $\guvb$ values generated with the semi-numerical method of \citet{davies_furlanetto_2016}. 
These boxes have a size $L_{\text{box}} = 512$ cMpc and $128^3$ pixels.
We have one snapshot of these $\guvb$ boxes at $z = 5.5$. 
To get the flux skewers used in this work, we combine random skewers of $\guvb$ from these UVB boxes with the skewers from the \texttt{Nyx} box.
The UVB boxes have a different resolution than the \texttt{Nyx} box, to generate a skewer of $\guvb$ values we randomly selected a starting location and direction in the UVB box then linearly interpolated the $\log(\guvb)$ values onto the same length and resolution as the \texttt{Nyx} skewers.

The method of \citet{davies_furlanetto_2016} allows for a spatially varying mean free path generated from fluctuations in the density of the sources of ionizing radiation with $\lambda \propto \guvb^{2/3} \Delta^{-1}$, for $\lambda$, the local mean free path, and $\Delta$, the local matter density. 
These simulations are scaled such that the mean value $\langle \lambda \rangle = \lammfp$ as desired. 
A brief summary of the \citet{davies_furlanetto_2016} method is as follows.
Cosmological initial conditions, independent of those from the 100 cMpc $h^{-1}$ \texttt{Nyx} boxes, were generated for the 512 cMpc box and evolved to z = 5.5 via the Zel’dovich approximation \citep{zel'dovich_1970}. 
Halos were created via the approach of \citet{mesinger_furlanetto_2007} down to a minimum halo mass of $M_{\text{min}} = 2 \times 10^9 M_{\odot}$. 
The ionizing luminosity of galaxies corresponding to each halo were determined following two steps: first the UV luminosities of galaxies were assigned by abundance matching to the \citet{bouwens_2015} UV luminosity function and then the ionizing luminosity of each galaxy was assumed to be proportional to its UV luminosity where the constant of proportionality is left as a free parameter.
The ionizing background radiation intensity, $J_{\nu}$, is then computed by a radiative transfer algorithm. 
The photoionization rate, $\guvb$, is finally calculated by integrating over $J_{\nu}$. 
For more details on the method see \citet{davies_furlanetto_2016}, \citet{davies_2018_abc} or Davies et al. 2022 in prep. where they also use this stitching procedure. 
Note that this method of generating UVB fluctuations ignores the effect of correlations between the baryon density in the \texttt{Nyx} boxes and the UVB. 
This is sufficient for the aims of this work but see Section \ref{section:mfp_density_match} for a discussion on the effects of ignoring these correlations on the resulting auto-correlation function and therefore future measurements of $\lammfp$ from real data.

Example slices through the UVB boxes for four values of $\lammfp$ are shown in Figure \ref{fig:mfp_slices} with a lower cutoff of $\log(\guvb / \langle\guvb\rangle) = -1$ for visual purposes.
The top left box shows a slice of $\guvb$ for the UVB simulation with the shortest $\lammfp = 5$ cMpc and has the greatest fluctuations. 
The bottom right box shows a slice of $\guvb$ for the UVB simulation with the longest $\lammfp = 150$ cMpc and has the weakest fluctuations. 
This follows since overall longer $\lambda$ values means that photons travel further and effectively smooth the UVB over these large scales. 

We ran UVB boxes for 14 values of $\lammfp$ (in cMpc): 5, 6, 8, 10, 15, 20, 25, 30, 40, 50, 60, 80, 100, and 150.
To generate UVB boxes for additional values of $\lammfp$ we linearly interpolated the $\log(\guvb)$ values at each location in the box between the two UVB boxes with the nearest $\lammfp$ values.
This was done for three linearly spaced values between each existing $\lammfp$ values, resulting in a total of 53 UVB boxes. 

To model a hypothetical evolution of $\lammfp$ as a function of redshift we used the double power law shown as the dot dashed line shown in Figure \ref{fig:mfp_model_evolution}. 
This double power law was fit by eye with the following two considerations. 
We fixed the low $z$ behavior to the power law fit from \citet{worseck_2014} for $z < 5$: $\lambda_{\text{mfp}}(z) = (37 \pm 2)h_{70}^{-1} [(1 + z) /5]^{-5.4 \pm 0.5}$ Mpc (proper).
We also required consistency with the new measurements at higher $z$ from \citet{becker_2021}. 
The resulting double power law is:
\begin{equation}
    \lambda_{\text{mfp}}(z) = \frac{37 h_{70}^{-1} \left(\frac{5}{6.55}\right)^{5.4}}{\left(\frac{1 + z}{6.55}\right)^{5.4} + \left(\frac{1 + z}{6.55}\right)^{25.5}} \text{ Mpc (proper).}
    \label{eq:double power law}
\end{equation}
We then evaluated equation \eqref{eq:double power law} at center of the seven redshift bins we considered and rounded to the nearest integer. 
The resulting true model $\lammfp$ values are listed in Table \ref{tab:central vals} and are plot as the black circles in Figure \ref{fig:mfp_model_evolution}. 
If these values were already in our set of 53 models then nothing else was done. 
If not, we linearly interpolated the value of $\log(\guvb)$ at each point in the UVB simulation box between the two UVB boxes with the closest values of $\lammfp$ to get the final desired UVB box. 
This ultimately caused some redshifts to have 53 models of $\lammfp$ while others have 54. 
To generate the final flux skewers, we calculated the optical depths assuming a constant UVB then rescaled $\tau_{\text{mfp}} = \tau_{\text{const.}}/(\guvb/\langle \guvb \rangle$). 
The $z = 5.5$ values of $\guvb$ are used when generating flux skewers at all redshifts. 
This is justified because the value of $\lammfp$ is more important than the redshift evolution of the bias of the source population between $5 \leq z \leq 6$ \citep{furnaletto_2017}.

\begin{table}
    \centering
    \caption{
    This table lists several relevant parameters for our simulations and mock data set. 
    The second column lists the ``true" values of the redshift-dependent $\lammfp$ calculated from equation \eqref{eq:double power law}.
    The third column gives the true values of $\langle F \rangle$ at each $z$ from \citet{bosman_2021_data}.
    These $\langle F \rangle$ values are the central value for the grid of values considered.
    The final column contains the number of quasar sightlines we modeled for one mock data set, which is the data set size in \citet{bosman_2021_data}.
    These sightlines each have a length of $\Delta z = 0.1$.
    }
	\label{tab:central vals}
        \begin{tabular}{|c|c|c|c|}
        \hline
        $z$ & $\lambda_{\text{mfp}}$ (cMpc) & $\langle F \rangle$ & \# QSOs \\ \hline
        5.4 & 39                            & 0.0801              & 64      \\
        5.5 & 32                            & 0.0591              & 64      \\
        5.6 & 26                            & 0.0447              & 59      \\
        5.7 & 20                            & 0.0256              & 51      \\
        5.8 & 16                            & 0.0172              & 45      \\
        5.9 & 12                            & 0.0114              & 28      \\
        6.0 & 9                             & 0.0089              & 19      \\ \hline
        \end{tabular}
\end{table}

The overall average of $\guvb$ calculated in the UVB fluctuation simulations is not uniquely determined since this originates from complicated galaxy physics. 
Thus, we force the average mean flux, $\langle F \rangle$, to be the same for each model where the average is taken over all flux skewers considered. 
This is achieved by calculating a constant, $a$, such that $\langle e^{-a\boldsymbol{\tau}} \rangle = \langle F \rangle$.
Additionally, we want to consider how changes in $\langle F \rangle$ would affect the auto-correlation function and determine if there is a degeneracy with $\lammfp$. 
Therefore we create a grid of 9 values of $\langle F \rangle$ at each redshift. 
We chose the central value of $\langle F \rangle$ for a grid from \citet{bosman_2021_data} and chose the range of values to keep $\langle F \rangle > 0$ while not running into boundary issues during our inference. 

\subsection{Comparison of Flux Skewers}

\begin{figure*}
	\includegraphics[width=2\columnwidth]{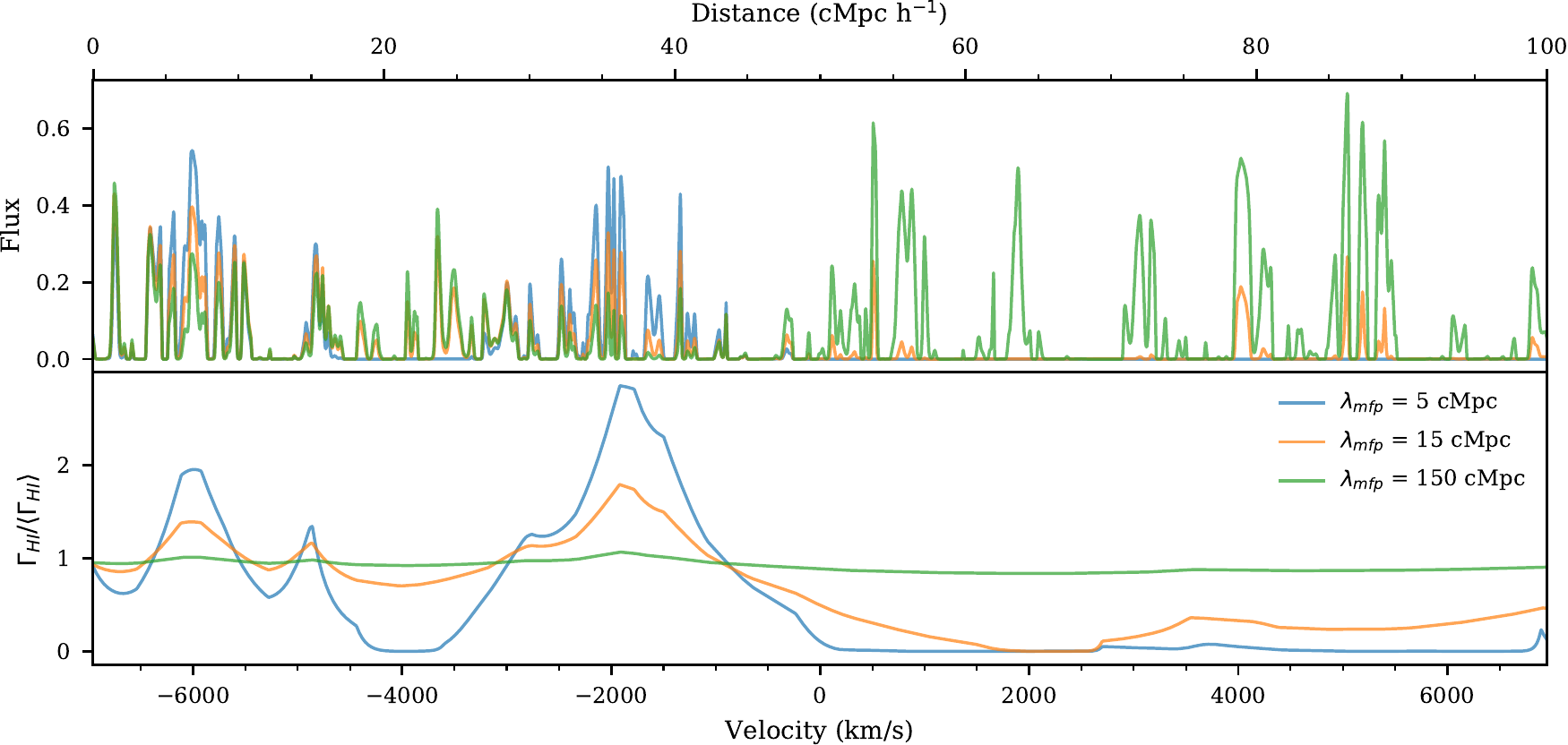}
    \caption{
        This figure shows the flux for one skewer of our simulation at at $z = 5.4$ with different values of $\lammfp$ all normalized to $\langle F \rangle = 0.0801$ in the top panel.
        The bottom panel shows the corresponding UVB skewer used to calculate the flux. 
        Smaller $\lammfp$ values (such as $\lammfp = 5$ cMpc in blue) has greater variations in $\guvb$ while the larger $\lammfp$ values (such as $\lammfp = 150$ cMpc in green) are more uniform. 
        Larger values of $\guvb$ leads to increased flux in that region which can be seen when comparing the two panels. 
        Consider $\Delta v = \SI{-2000}{\kilo\meter\per\second}$, here $\lammfp = 5$ cMpc model (blue) has a peak in $\guvb$. 
        The corresponding flux is boosted when compared to the other models. 
        Additionally, for $\lammfp = 5$ cMpc (blue) the $\guvb$ values are very small for $\Delta v \geq \SI{0}{\kilo\meter\per\second}$ resulting in $F \sim 0$. 
    }
    \label{fig:flux_three_paenl}
\end{figure*}

We drew 1000 skewers from the \texttt{Nyx} simulation and 1000 independent skewers of $\guvb$ from the UVB boxes to use in this work.
One example flux skewer, which combines the \texttt{Nyx} simulation skewer and the $\guvb$ values from the UVB boxes, at $z = 5.4$ is shown in Figure \ref{fig:flux_three_paenl} for three different values of $\lammfp$ all normalized to $\langle F \rangle = 0.0801$.
The bottom panel of this figure shows the corresponding UVB skewers that were used to calculate the flux.
2D slices of the UVB boxes these skewers came from are shown in Figure \ref{fig:mfp_slices}. 
The model shown with the shortest $\lammfp$, 5 cMpc (blue), results in the greatest variation of $\guvb/\langle\guvb\rangle$. 
In particular, note that at $\Delta v = \SI{-2000}{\kilo\meter\per\second}$, the $\lammfp = 5$ cMpc model (blue) has a peak in $\guvb$ and the corresponding flux is boosted when compared to the other models. 
Additionally, for $\lammfp = 5$ cMpc (blue) the $\guvb$ values are very small for $\Delta v \geq \SI{0}{\kilo\meter\per\second}$ resulting in $F \sim 0$. 
The model with the largest $\lammfp$, 150 cMpc (green), shows a mostly uniform $\guvb$ skewer throughout the whole velocity range leading to more consistent flux levels.

\subsection{Forward Modeling}

For this work we aim to model the resolution, noise, and size properties of a realistic data set. 
We first chose to model a simplified version of the XQR-30 (main and extended) data set\footnote{https://xqr30.inaf.it/}. 
The main XQR-30 data set consists of 30 spectra of the brightest $z > 5.8$ quasars observed with VLT/X-shooter \citep{vernet_2011_xshooter}.
These spectra are supplemented with an extended data set consisting of 12 archival X-shooter spectra with comparable signal-to-noise ratio. 
See D'Odorico in prep. for additional information on these data. 
For this work we specifically model properties similar to the data set of \citet{bosman_2021_data} which consists of the etended XQR-30 data supplemented with additional archival X-Shooter data and archival Keck/ESI spectra which have a lower resolution than the X-shooter spectra.

For our simplified modeling, we use the resolving power of X-shooter for visible light with a 0.9" slit, so $R = 8800$. 
We also use a typical signal to noise ratio per \SI{10}{\kilo\meter\per\second} pixel ($\snr$) of $\snr = 35.9$, which is the median of all the data presented in \citet{bosman_2021_data}. 
Additionally, we investigate how higher resolution data with access to smaller scales in the \lya forest would impact measurements of $\lammfp$ from the auto-correlation function. 
To achieve this we consider a ``high-resolution" data set with the same $\snr$ and size properties as the ``low-resolution" ($R = 8800$) data set but with $R = 30000$. 
This resolution is achievable with instruments such as Keck/HIRES, VLT/UVES, and Magellan/MIKE though the number of sightlines and noise properties used here do not represent a high-resolution data set currently in existence. 

We model the resolution by smoothing the flux by a Gaussian filter then after smoothing we re-sampled such that there are 4 pixels per resolution element, where the resolution element is the FWHM. 
This means, for the low-resolution data set we smoothed by a Gaussian filter with $\text{FWHM} \approx \SI{34}{\kilo\meter\per\second}$ then re-sampled so the pixel size was $\Delta v = \SI{8.53}{\kilo\meter\per\second}$.
For the high-resolution data set we smoothed by a Gaussian filter with $\text{FWHM} = \SI{10}{\kilo\meter\per\second}$ then re-sampled so the pixel size was $\Delta v = \SI{2.5}{\kilo\meter\per\second}$.

As stated above, we modeled a $\snr = 35.9$. 
Using $\text{SNR}_{\Delta v} = \snr \sqrt{\Delta v / \SI{10}{\kilo\meter\per\second}}$ this corresponds to a signal to noise ratio of 33.2 per 8.53 km/s low-resolution pixel and a signal to noise ratio of 18.0 per 2.5 km/s high-resolution pixel.
For simplicity, we add flux-independent noise in the following way.
We generate one realization of random noise drawn from a Gaussian with $\sigma_N = 1 / \text{SNR}_{\Delta v}$ for each SNR value and add this noise realization to every model at every redshift.
The size of each noise realization is the number of skewers created (1000) by the number of pixels in the re-sampled flux skewers (1705 pixels for low-resolution and 5814 pixels for high-resolution).
Using the same noise realization over the different models prevents stochasticity from different realizations of the noise from causing a noisy likelihood, which means the likelihood will be smooth as a function of model parameter. 
Thus the noise modeling will not unduly, adversely effect the parameter inference.

\begin{figure}
	\includegraphics[width=\columnwidth]{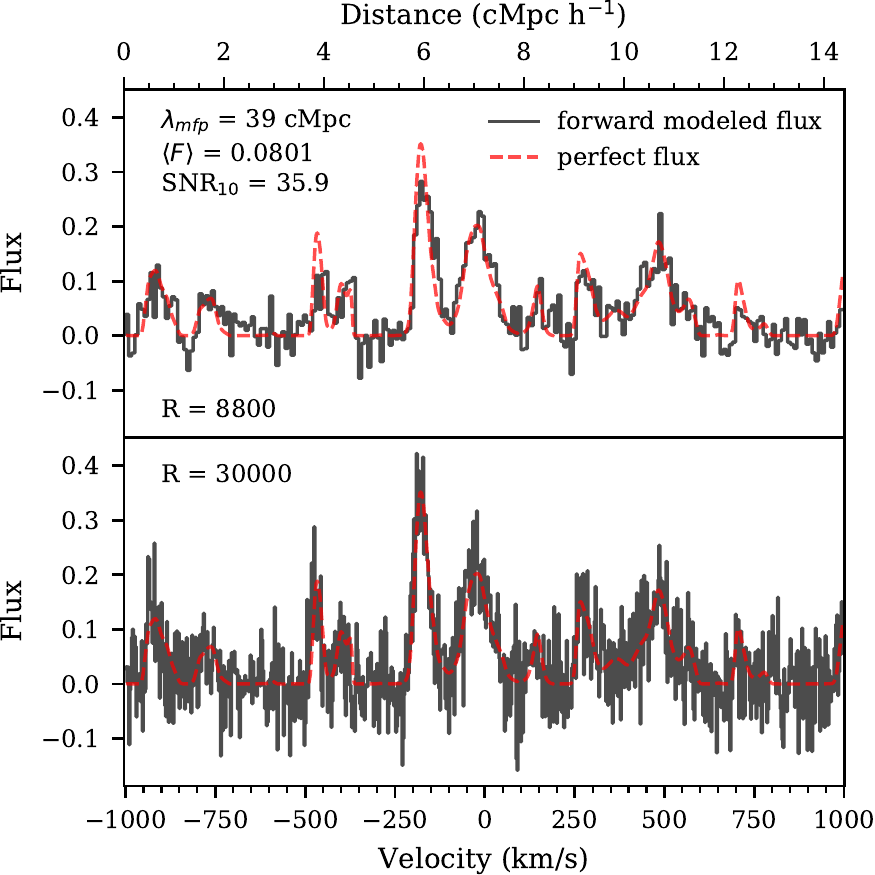}
    \caption{
        Both panels show initial and forward-modeled flux from a skewer with $\lammfp = 39$ cMpc and $\langle F \rangle = 0.0801$ at $z = 5.4$.
        The initial flux is the same in both panels (red dashed line) while the forward modeled flux (black histogram) varies. 
        The top panel shows the low-resolution flux with $R = 8800$, which represents XQR-30 data. 
        The bottom panel shows the high-resolution flux with $R = 30000$.
        Both of these resolutions have $\snr = 35.9$ which leads to differing $\text{SNR}_{\Delta v}$ as can be seen when comparing the two panels. 
    }
    \label{fig:flux_noise}
\end{figure}

A section of one skewer for both the initial and forward-modeled flux is shown in Figure \ref{fig:flux_noise}. 
Both panels shows a skewer at $z = 5.4$ with $\lammfp = 39$\ cMpc and $\langle F \rangle = 0.0801$, our assumed true parameter values at this redshift. 
The initial flux in both panels is the same and is shown as a red dashed line. 
The top panel shows the low-resolution forward-modeled flux (black histogram) with 
$R = 8800$.
The bottom panel shows the high-resolution forward-modeled flux (black histogram) with $R = 30000$. 
Again both of these panels have the same $\snr = 35.9$ which results in different noise levels per pixel, as can be seen when comparing the two panels.

We assume a fiducial data set size that matches the number of sightlines reported in Table 4 of \citet{bosman_2021_data} each with a length of $\Delta z = 0.1$. 
The number of sightlines are reported in the last column of Table \ref{tab:central vals} where to total pathlength considered is equal to these values multiplied by $\Delta z = 0.1$. 
Redshift bins of $\Delta z = 0.1$ correspond to distances of 33 to 29 cMpc $h^{-1}$ when centered at $z = 5.4$ to $z = 6.0$. 
However, the \texttt{Nyx} simulation box is 100 cMpc $h^{-1}$ long, much longer than these redshift bins. 
If we were to use the full 100 cMpc $h^{-1}$ skewers in our calculation we would be averaging over fewer skewers to get the same total $\Delta z$ path.
We wanted to use a greater number independent skewers with more accurate lengths when compared with observed \lya forest regions.
For simplicity, we split all our skewers into two 40 cMpc $h^{-1}$ regions which we treated as independent, giving us an effective number of 2000 independent skewers. 

Note that unless otherwise specified the plots in this work mainly show results from the low resolution, $R = 8800$ data, since it represents existing XQR-30 data.

\section{Methods} \label{section: methods}

\subsection{Auto-correlation function} \label{section: autocorr}

The auto-correlation function of the flux ($\xi_F (\Delta v)$) is defined as
\begin{equation}
    \xi_F (\Delta v) = \langle F(v) F(v + \Delta v) \rangle
\end{equation}
where $F(v)$ is the flux of the \lya forest and the average is performed over all pairs of pixels at the same velocity lag ($\Delta v$). 
The auto-correlation function is related to the power spectral density ($P_F(k)$) as 
\begin{equation}
    P_F(k) = \langle F \rangle^{-2} \int_{-\infty}^{\infty} \xi_F (\Delta v) e^{-i k \Delta v} d(\Delta v).
    \label{eq:power ft}
\end{equation}
Note that this implies that the auto-correlation function should be sensitive to the same physical parameters as the power spectrum. 
Additionally, the auto-correlation function has nice properties with respect to white noise and spectral masks that make it a promising statistic to measure. 
Conventionally, the flux contrast field, $(F - \langle F \rangle) / \langle F \rangle$, is used when measuring statistics of the \lya forest. 
Here, we chose to use the flux since $\langle F \rangle$ is small and has large uncertainties at high-$z$ where we are most interested in this measurement.
Using the flux thus prevents us from dividing by a small number which would come from an independent measurement and could potentially blow up the value of the flux contrast. 
This leads to the factor of $\langle F \rangle ^{-2}$ in Equation \eqref{eq:power ft}.

For each resolution and model we compute the auto-correlation function with a bin size of one FWHM of the resolution (either $\SI{34}{\kilo\meter\per\second}$ or $\SI{10}{\kilo\meter\per\second}$) starting from this resolution size out to 20 cMpc $h^{-1}$ (half the length of the skewer) which corresponds to $\sim \SI{2900}{\kilo\meter\per\second}$ at $z = 5.4$. 
The model value of the auto-correlation function was determined by taking the average of the auto-correlation function over all 2000 forward-modeled skewers.
Each mock data set of the auto-correlation were calculated by taking an average over the appropriate number of random skewers for the number of quasars at that redshift from the initial 2000 forward-modeled skewers.
The value of the auto-correlation function for small-scale bins is affected by the finite resolution. This effect is left in both the models and the mock data. 
We determine the errors on the models via the following estimate of the covariance matrix from mock draws of the data:
\begin{equation}
    \Sigma(\boldsymbol{\xi_\text{model}}) = \frac{1}{N_{\text{mocks}}} \sum_{i=1}^{N_{\text{mocks}}}(\boldsymbol{\xi}_i - \boldsymbol{\xi_\text{model}})(\boldsymbol{\xi}_i - \boldsymbol{\xi_\text{model}})^{\text{T}}
    \label{eq:covariance}
\end{equation}
where $\boldsymbol{\xi}_i$ is the auto-correlation function calculated for the i-th mock data set, $\boldsymbol{\xi_\text{model}}$ is the average value of the auto-correlation function over all 2000 skewers, and $N_{\text{mocks}}$ is the number of forward-modeled mock data sets used.
Both the mock data sets and the overall average have the same values of $\lammfp$ and $\langle F \rangle$ in this calculation, so we end up with a covariance matrix at each parameter grid point.
We use $N_{\text{mocks}} = 500000$ for all models and redshifts in this work, see Appendix \ref{appendix:converge} for a discussion on the convergence of the covariance matrix.

\begin{figure*}
	\includegraphics[width=2\columnwidth]{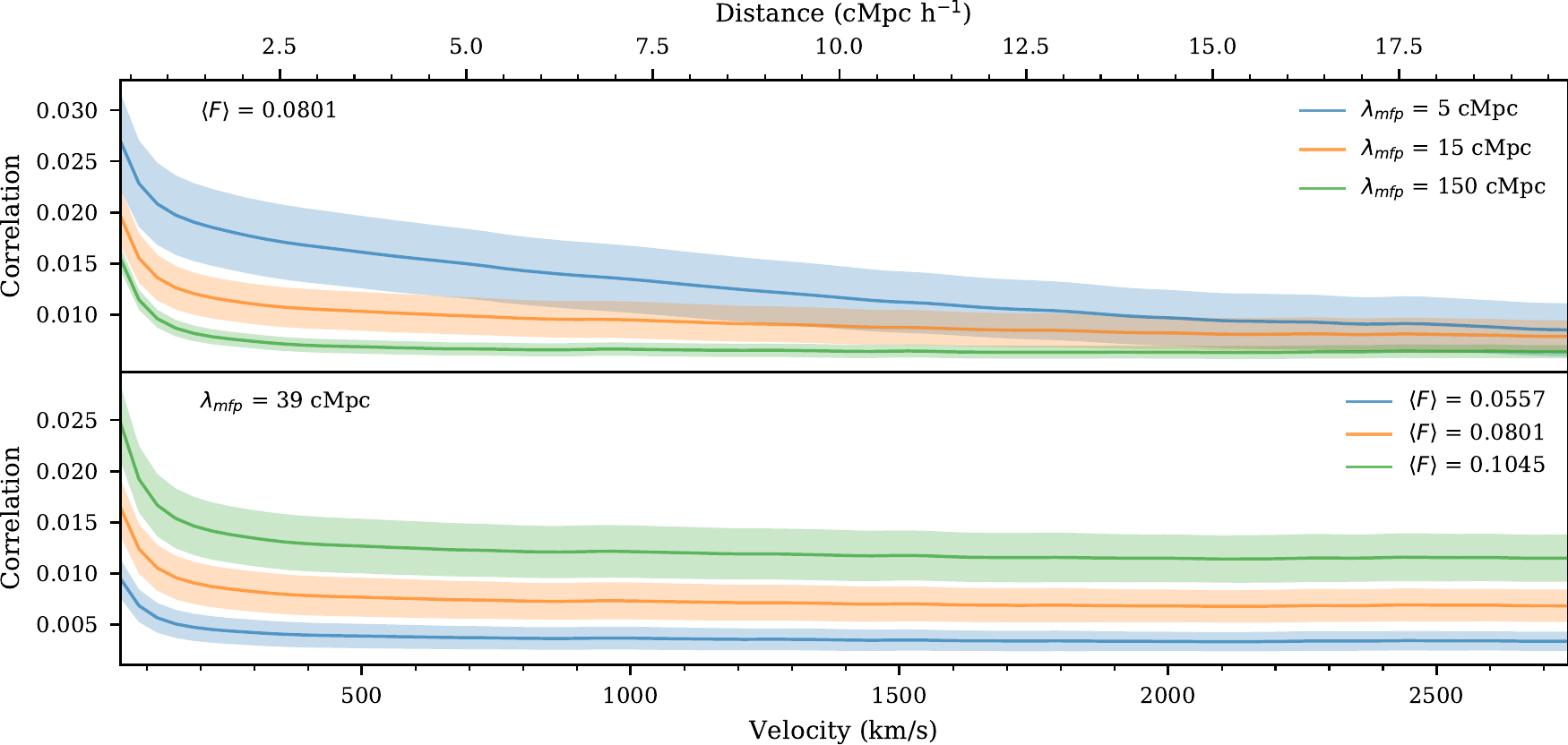}
    \caption{
        This figure demonstrates the effects of varying $\lammfp$ and $\langle F \rangle$ on the model values of the auto-correlation function at $z = 5.4$ and $R = 8800$.
        The solid lines show the model values calculated by averaging the auto-correlation function from all forward modeled skewers available while the shaded regions show the errors from the covariance matrix as estimated in equation \eqref{eq:covariance}. 
        The top panel varies $\lammfp$ with a constant $\langle F \rangle$ labeled in the top left corner while the bottom panel does the opposite.
        Both $\lammfp$ and $\langle F \rangle$ change the auto-correlation function on all scales shown, though $\lammfp$ appears to effect small scales more than large scales. 
        In the top panel, the model value of the auto-correlation function are further apart for $\lammfp = 5$ cMpc (blue) and $\lammfp = 15$ cMpc (orange) than for $\lammfp = 15$ cMpc (orange) and $\lammfp = 150$ cMpc (green), which is a greater difference in $\lammfp$ value. 
        This means the auto-correlation function is more sensitive to small $\lammfp$ values than large $\lammfp$ values. 
        Comparatively, in the bottom panel, the differences in the mean auto-correlation function appear roughly linear with varying $\langle F \rangle$ which should result in similar sensitivity for all $\langle F \rangle$ values. 
    }
    \label{fig:correlation_two_panels}
\end{figure*}

Figure \ref{fig:correlation_two_panels} shows the model value of the auto-correlation function with different parameter values at $z = 5.4$. 
The top panel shows models with a changing $\lammfp$ and constant $\langle F \rangle = 0.0801$. 
The solid lines show the model values calculated by averaging the auto-correlation function from all forward modeled skewers while the shaded regions show the errors from the diagonal elements of the covariance matrix as estimated in equation \eqref{eq:covariance}. 
Smaller $\lammfp$ values (such as $\lammfp = 5$ cMpc - blue) result in a greater 
correlation function at all scales, though mainly at small scales, and larger error bars than large $\lammfp$ values (such as $\lammfp = 150$ cMpc - green). 
These models are non-linearly spaced with greater differences between the models at small $\lammfp$ (blue vs orange) than large $\lammfp$ (orange vs green) which will result in variable sensitivity to $\lammfp$ from the auto-correlation function at different $\lammfp$ values.
The bottom panel shows models with varying $\langle F \rangle$ and constant $\lammfp = 39$ cMpc.
$\langle F \rangle$ sets the overall amplitude of the auto-correlation function.
Here the differences between models are linear where larger $\langle F \rangle$ leads to larger auto-correlation values. 
This scaling is roughly $\propto \langle F \rangle^2$, which follows from the definition of the auto-correlation function. 

\begin{figure}
	\includegraphics[width=\columnwidth]{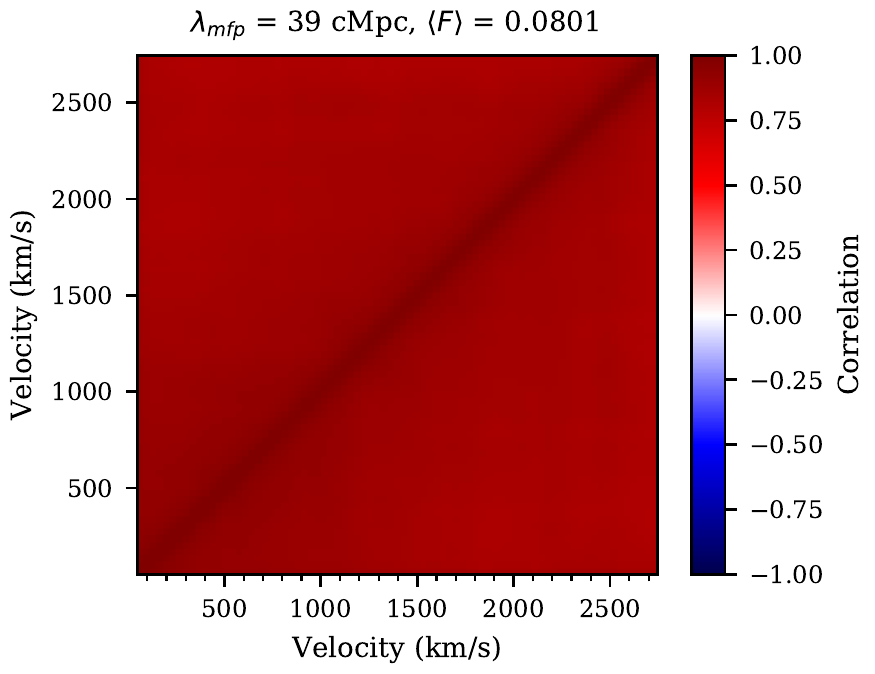}
    \caption{
        This figure shows the correlation matrix calculated with equation \eqref{eq:correlation} with $N_{\text{mocks}} = 500000$ for the model at $z = 5.4$ with $\lammfp = 39$ cMpc, $\langle F \rangle = 0.0801$, and $R = 8800$.
        The color bar is fixed to span from -1 to 1, which is all possible values of the correlation matrix. 
        Here it is clear that all bins in the auto-correlation function are highly correlated with each other. 
    }
    \label{fig:covar_correlation}
\end{figure}

To visualize the covariance matrix, we define the correlation matrix, $C$. 
The correlation matrix is the covariance matrix with the diagonal normalized to 1. 
This is done to the $j$th, $k$th element by
\begin{equation}
        C_{jk} = \frac{\Sigma_{jk}}{\sqrt{\Sigma_{jj}\Sigma_{kk}}}.
        \label{eq:correlation}
\end{equation}
One example correlation matrix is shown in Figure \ref{fig:covar_correlation} for $z = 5.4$, $\lammfp$ $= 39$ cMpc, and $\langle F \rangle = 0.0801$.
All bins of the auto-correlation function are very-highly correlated which is due to the fact that each pixel in the \lya forest contribute to multiple (in fact almost all) bins in the auto-correlation function. 

\subsection{Effect of Model Limits on the Auto-correlation Function} \label{section:mfp_density_match}

As stated in Section \ref{section: sim data}, the semi-numerical method to generate the fluctuating UVB with various $\lammfp$ ignores the correlation between the density and $\guvb$. 
This is a result of the current limitations on available simulation boxes.
We require that the UVB boxes are large enough to avoid suppressing UVB fluctuations and we require that the underlying hydrodrynamical simulation boxes of the IGM have a grid that is fine enough to resolve the small structures in the \lya forest. 
\citet{lukic_2015} found that this grid needs to have a grid resolution of 20 $h^{-1}$ kpc to produce 1\% convergence of \lya forest flux statistics. 
\citet{davies_furlanetto_2016} found that, with their 400 Mpc box of $\guvb$ values, the tail of their optical depth distribution was impacted by cosmic variance, highlighting the need to go to even larger boxes. 
Having both a large box with a fine grid, which would be required to correlate the UVB and simulation box density, is currently too computationally expensive to be feasible.

\begin{figure*}
	\includegraphics[width=2\columnwidth]{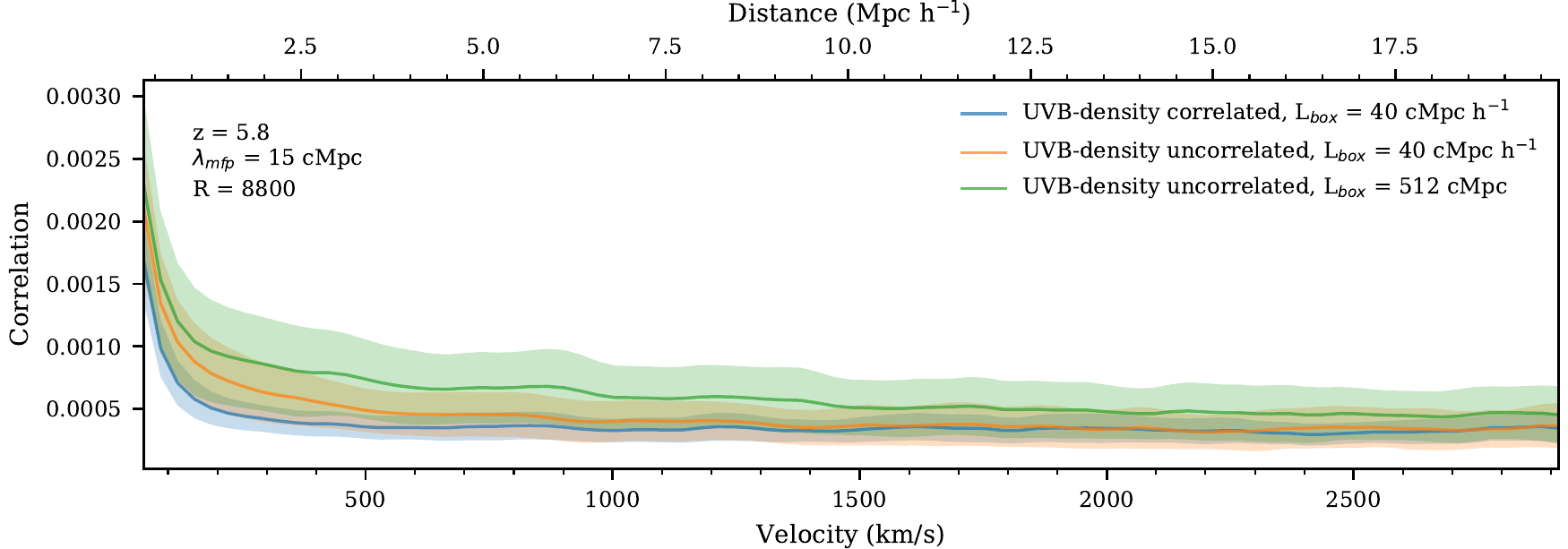}
    \caption{
        This figure demonstrates the effect of ignoring density correlations as well as using a small box size when generating $\guvb$. 
        The blue line shows the auto-correlation function when using a $\guvb$ calculated with the appropriate density field and a box size of $L_{\text{box}} = 40$ cMpc $h^{-1}$. 
        The orange line shows the same for a $\guvb$ calculated with the a random density field and a box size of $L_{\text{box}} = 40$ cMpc $h^{-1}$, isolating the effect of density correlations when compared to blue. 
        The green line shows the same for a $\guvb$ calculated with the a random density field and a box size of $L_{\text{box}} = 512$ cMpc, isolating the effect of the box size when compared with orange.
        Here we see that the correct density field will cause the signal on small scales to be reduced and that using a larger box size will increase the signal for all scales. 
    }
    \label{fig:uvb_box_size_orientation}
\end{figure*}

In general, there is a positive correlation between density and $\guvb$ and a negative correlation between density and transmitted flux. 
This means that in areas with high $\guvb$ there should also be higher density which would in turn decrease the transmitted flux, therefore reducing the extra signal from the short $\lammfp$.
To quantitatively explore this, we used a \texttt{Nyx} simulation box with a size of $L_{\text{box}} = 40$ cMpc $h^{-1}$ at $z = 5.8$. 
This box size has associated UVB values for $\lammfp = 15$ cMpc generated with the same method of \citet{davies_furlanetto_2016} as described in Section \ref{section: sim data uvb model}. 
For these UVB boxes the local matter density matches that of the \texttt{Nyx} simulations of the IGM.
We selected skewers from the UVB boxes in two ways: from the same location as the \texttt{Nyx} skewers or from a random location in the box.
When the UVB skewers come from the same location as the \texttt{Nyx} skewers the density field and UVB field are correlated. 
When the UVB skewers come from a random location these two fields will not be correlated, which is analogous to the uncorrelated modeling adopted in the main text. 
The resulting auto-correlation function models are shown in Figure \ref{fig:uvb_box_size_orientation} as the blue and orange lines. 
The blue line is the model with UVB skewers from the 40 $h^{-1}$ cMpc box that were derived from the same density field as the \texttt{Nyx} simulations.
The orange line is the model with UVB skewers from the 40 $h^{-1}$ cMpc box that were derived from a random density field.
Comparing these two lines isolates the effects of ignoring the UVB-density correlations. 
Here we see that the density correlations reduce the auto-correlation signal at small scales while leaving the large scale signal unchanged. 
When correlated, $\guvb$ is proportional to the local density field so the regions of high $\guvb$ values will also be regions of higher density. 
Since the optical depth scales as a power of the local density field, the boosted signal on small scales from regions of high $\guvb$ in the orange model will be reduced by the corresponding increased local density leading to the reduction in small scales in the blue model. 
Since the reduction is happening on small scales, this mimics the effect of instead having a model with a larger $\lammfp$.

Additionally, we investigated the effect of the box size used to generate the UVB on the amount of fluctuations in $\guvb$ seen at a fixed $\lammfp$. 
Using a smaller box size, such as the 40 $h^{-1}$ cMpc box considered in \citet{onorbe_2019}, can suppress fluctuations in the local $\lambda$ value since there is a smaller volume that must average to $\lammfp$. 
For this comparison, we use randomly selected UVB skewers from the 40 $h^{-1}$ cMpc box as well as randomly selected skewers from our 512 cMpc UVB box with $\lammfp = 15$ cMpc from the main text of this work as described in Section \ref{section: sim data uvb model}. 
The UVB skewers chosen with both of these methods are uncorrelated with the density field, so we isolate the effect of only the box size. 
The two resulting auto-correlation function models are also shown in Figure \ref{fig:uvb_box_size_orientation}. 
Again, the orange line is the model with UVB skewers from the 40 $h^{-1}$ cMpc box that were derived from a random density field.
The green line shows the  model with UVB skewers from the 512 cMpc box that has a random density field compared to the \texttt{Nyx} simulation. 
Comparing this green line to the orange line thus isolates the effect of the small box size where again the large box size is required for UVB fluctuations to converge for a given $\lammfp$. 
Here we see that the green model has a greater signal than the orange at all scales.
Therefore, both the blue and orange models with UVB skewers generated in a 40 $h^{-1}$ cMpc box are likely underestimating the auto-correlation function on all scales. 
This makes it difficult to quantify the level of excess signal in the auto-correlation function that we get from ignoring correlations between the UVB and the local density field since the signal is underestimated on all scales when using the smaller UVB box. 
For this reason, we choose not to correct the mock data to account for the effect of using an uncorrelated UVB.

The mock data and models of the auto-correlation function from this study are self-consistently generated since both ignore the correlations between the UVB and the local density field. 
Therefore, the excess signal on small scales from modeling with an uncorrelated UVB will not bias the constraints obtained in this work. 
However, this excess on small scales needs to be accounted for when using the models of the auto-correlation function from the main text to constrain $\lammfp$ with observational data. 
We would expect measurements made by comparing data to models generated without UVB-density correlations to be biased towards larger values of $\lammfp$, since the reduced signal on small scales from real density correlations would look like a larger $\lammfp$ in our models.
We can not quantify this potential bias with these simulations because, again, the small box size of 40 $h^{-1}$ cMpc reduces the auto-correlation function signal on all scales. 
Modeling the UVB consistently with Ly$\alpha$ forest simulations in larger boxes is necessary to conclusively study the limitations of the model used in this work. 
We therefore leave this to future work.

\subsection{Parameter Estimation} \label{section: mcmc}

To quantify the precision with which $\lammfp$ can be measured we use Bayesian inference with a multi-variate Gaussian likelihood and a flat prior over the parameters of interest.
This likelihood ($\mathcal{L} = p(\boldsymbol{\xi} | \lammfp, \langle F \rangle)$) has the form:
\begin{equation}
        \mathcal{L} = \frac{1}{\sqrt{\det(\Sigma) (2 \pi)^{n}}} \exp \left( -\frac{1}{2} (\boldsymbol{\xi} - \boldsymbol{\xi_\text{model}})^{\text{T}} \Sigma^{-1} (\boldsymbol{\xi} - \boldsymbol{\xi_\text{model}}) \right) 
        \label{eq:gauss_like}
\end{equation}
where $\boldsymbol{\xi}$ is the auto-correlation function from our mock data, $\Sigma = \Sigma(\boldsymbol{\xi_\text{model}})$ is the model dependent covariance matrix estimated by equation \eqref{eq:covariance}, and $n$ is the number of points in the auto-correlation function. 
We discuss the assumption of using a multivariate Gaussian likelihood in Appendix \ref{appendix:multi gaussian}. 

Our models are defined by two parameters: $\lammfp$ and $\langle F \rangle$.
We  compute the posteriors for these parameters using Markov Chain Monte Carlo (MCMC) with the \texttt{EMCEE} package \citep{forman_mackey_2013}.
We linearly interpolate the model values and covariance matrix elements onto a finer 2D grid of $\lammfp$ and $\langle F \rangle$ then use the nearest model during the MCMC. 
This fine grid has 137 values of $\lammfp$ and 37 values of $\langle F \rangle$.
Our MCMC was run with 16 walkers taking 5000 steps each and skipping the first 500 steps of each walker as a burn-in. 

\begin{figure*}
    \centering
    
    \begin{subfigure}[t]{\textwidth}
    \centering
        \includegraphics[width=\linewidth]{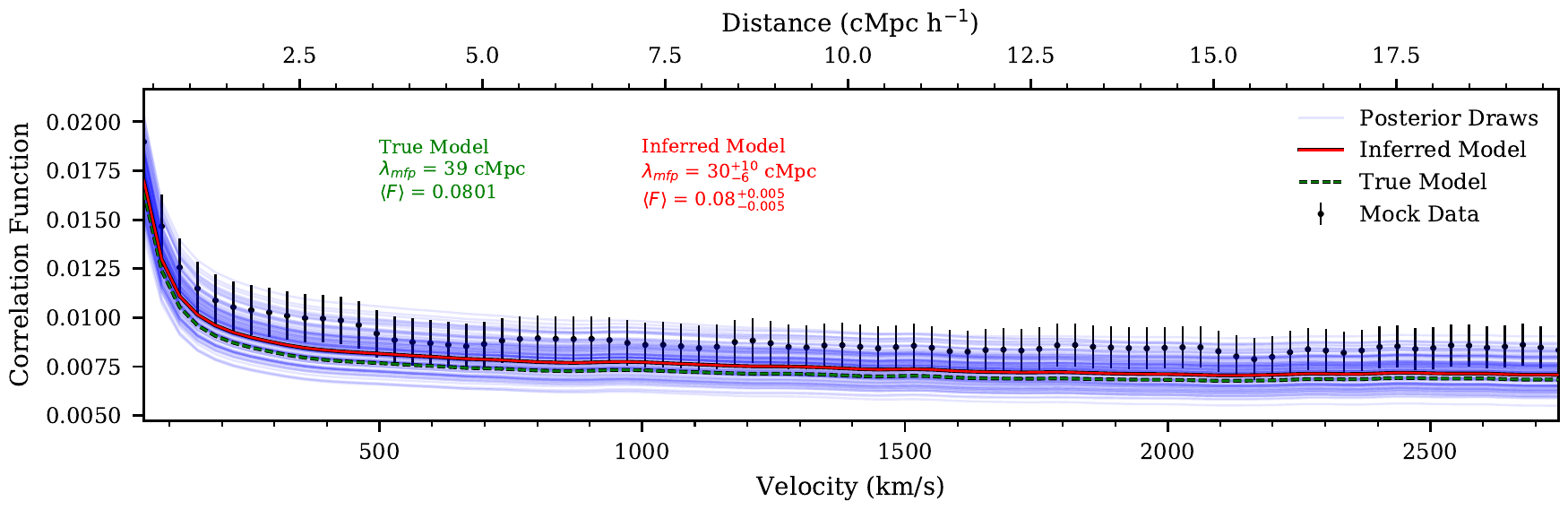} 
    \end{subfigure}
    
    \vspace{3mm}

    \begin{subfigure}[t]{0.49\textwidth}
        \centering
        \includegraphics[width=\linewidth]{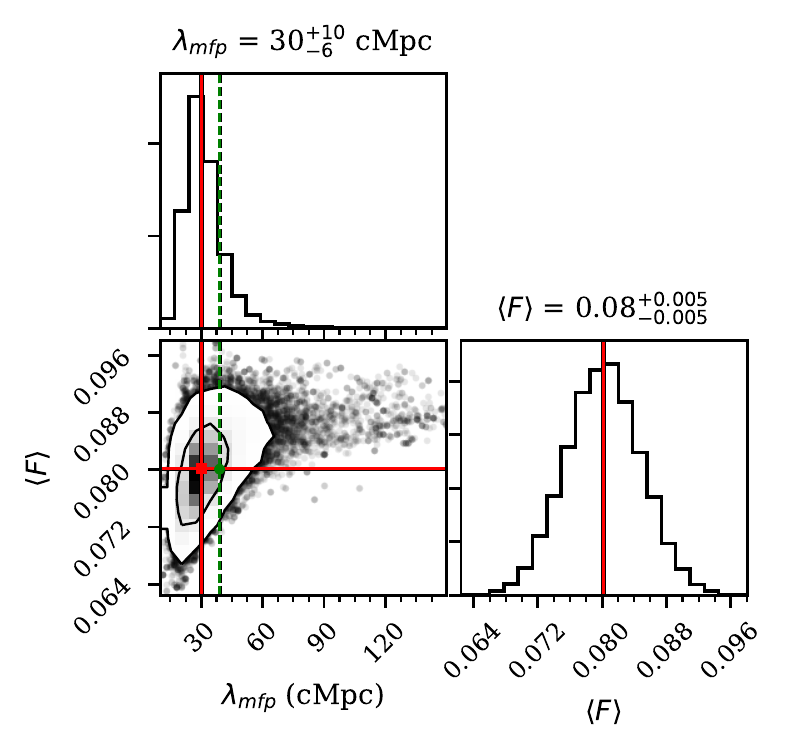} 
    \end{subfigure}
    \hfill
    \begin{subfigure}[t]{0.49\textwidth}
        \centering
        \includegraphics[width=\linewidth]{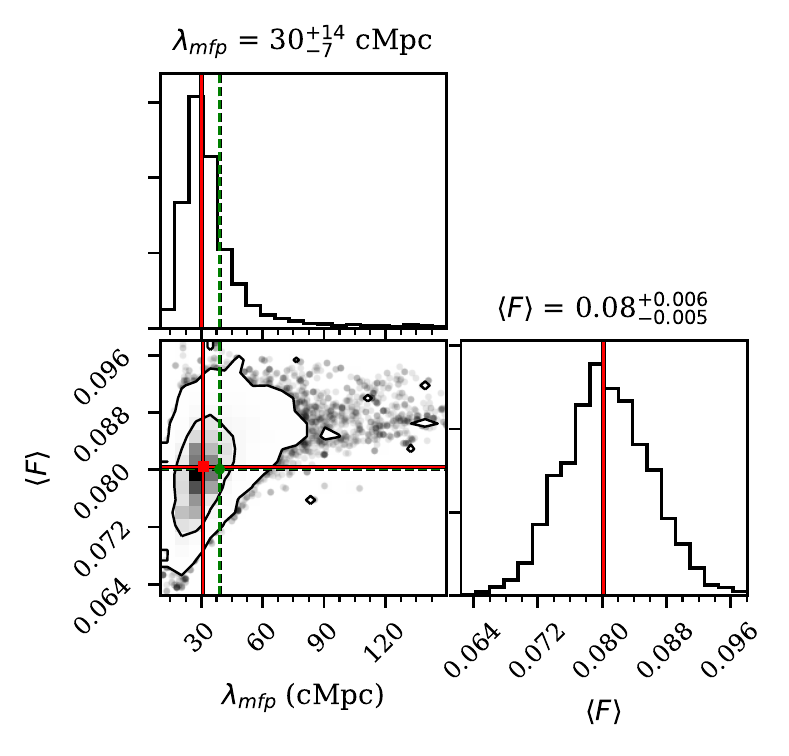} 
    \end{subfigure}
    
    \caption{
        This figure illustrated the results of our inference procedure applied to one mock data set at $z = 5.4$. 
        The top panel shows the resulting models from our inference procedure without re-weighting while the bottom panel has two corner plots that show the resulting parameters, the left without re-weighting and the right with re-weighting. 
        In the top panel, the black points with error bars are the mock data with error bars from the inferred model. 
        The inferred model was calculated by the median (50th percentile) of the MCMC chains of each parameter independently. 
        The inferred model is shown as a red line while the accompanying red text reports errors calculated from the 16th and 84th percentiles of each parameter.
        In comparison, the true model the data was drawn from is the green dotted line and accompanying text. 
        To demonstrate the width of the posterior, multiple faint blue lines are shown which are the models corresponding to the parameters from 100 random draws of the MCMC chain. 
        The bottom left panel shows a corner plot of the values of $\lammfp$ and $\langle F \rangle$ that immediately result from our inference procedure. 
        The bottom right panel shows the corner plot of the values of $\lammfp$ and $\langle F \rangle$ from our inference procedure using the re-weighting approach. 
        This means the corner plot has been made with the weights calculated from our inference test as described in Section \ref{section: inf re-weight}
    }
    \label{fig:corner_fit}
\end{figure*}

Figure \ref{fig:corner_fit} shows the result of our inference procedure for one mock data set at $z = 5.4$. 
The top panel shows the mock data set with various lines relating to the inference procedure as follows. 
The green dotted line and accompanying text presents the true model that the mock data was drawn from. 
The mock data set is plot as the black point with error bars that come from the diagonal elements of the covariance matrix of the model that is nearest to the inferred model.
The inferred model is the model that comes from the median of each parameter determined independently via the 50th percentile of the MCMC chains. 
The red lines and accompanying text shows the inferred model from MCMC. 
The errors on the inferred model written in the text are the distance between the 16th, 50th, and 84th percentiles of the MCMC chains. 
The blue lines show the models corresponding to 100 random draws from the MCMC chain to visually demonstrate variety of models that come from the resulting posterior. 
The bottom left panel shows a corner plot of the posteriors for both $\lammfp$ and $\langle F \rangle$.
Here we see evidence of an extended tail out towards larger $\lammfp$ which is quantified in the asymmetric errors reported on the inferred value of $\lammfp$. 
This asymmetry comes from the non-linear spacing of the auto-correlation function models as discussed in Section \ref{section: autocorr}.

\subsection{Inference Test and Re-weighting} \label{section: inf re-weight}

We perform a test to check the fidelity of our inference procedure in order to ensure that our resulting posteriors act in a statistically valid way.
This will ensure any assumptions we make during our inference are justified. 
For example, in this work we used an approximate likelihood in the form of a multivariate Gaussian likelihood.
The \lya forest is known to be a non-Gaussian random field. 
By adopting a multivariate Gaussian likelihood here, we are tacitly assuming that averaging over all pixel pairs when calculating the auto-correlation function will Gaussianize the resulting distribution of the values of the auto-correlation function, as is expected from the central limit theorem. 
We discuss the distribution of these values for our mock data in detail in Appendix \ref{appendix:multi gaussian}. 
If this assumption is not valid our reported errors may be either underestimated or overestimated.

The general idea of our inference test is to compare the true probability contour levels with the ``coverage" probability. 
The coverage probability is the percent of time the probability of the true parameters of a mock data set fall above a given probability level over many mock data sets. 
In our case, we compute this over 500 mock data sets where the true parameters considered are samples from our priors. 
Ideally, this coverage probability should be equal to the chosen probability contour level. 
This calculation can be done at many chosen probabilities resulting in multiple corresponding coverage probabilities.
Existing work that explore this coverage probability include \citet{prangle_2014, ziegel_2014, morrison_2018, sellentin_2019}.

\begin{figure*}
    \centering
    \begin{subfigure}[t]{0.49\textwidth}
        \centering
        \includegraphics[width=\linewidth]{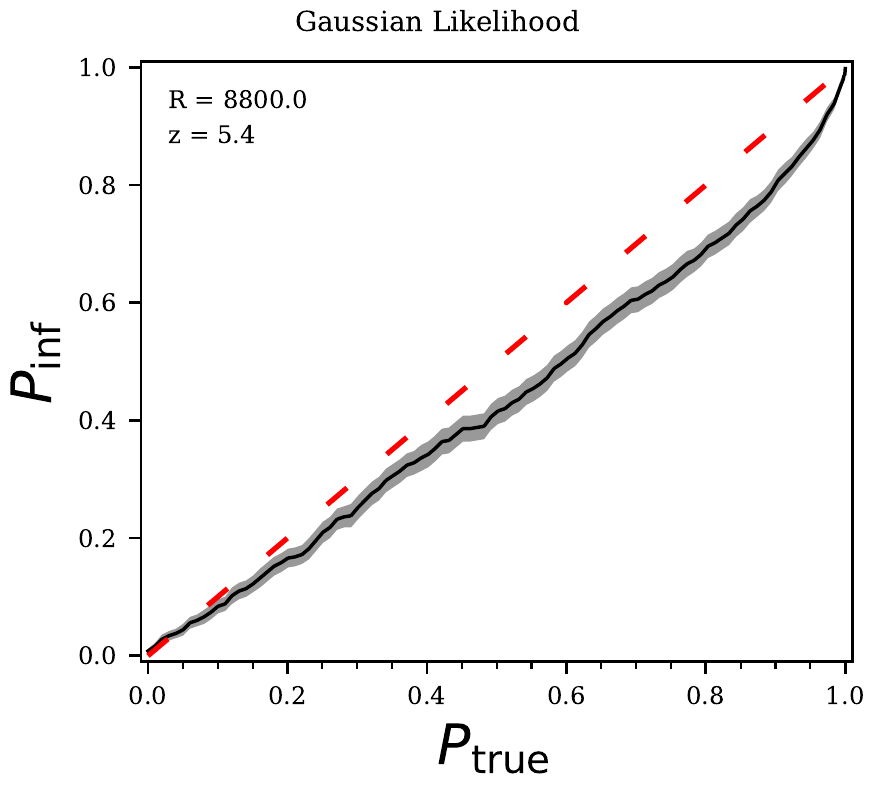} 
    \end{subfigure}
    \hfill
    \begin{subfigure}[t]{0.49\textwidth}
        \centering
        \includegraphics[width=\linewidth]{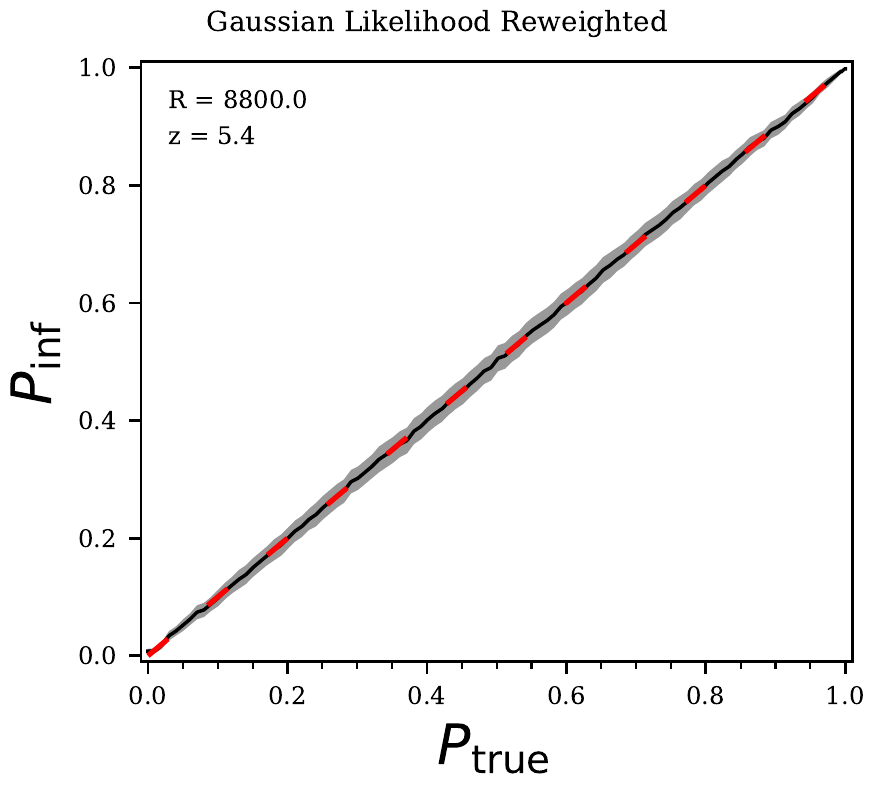} 
    \end{subfigure}
    \caption{
        The left panel of this figure shows the coverage resulting from the inference test from 500 models at $z = 5.4$ and $R = 8800$ drawn from our priors on $\lammfp$ and $\langle F \rangle$.
        Here we see that the true parameters for the models fall above the 60th percentile in the MCMC chain $\sim 50\%$ of the time, for example. 
        The right panel of this figure shows the coverage resulting from the inference test with the use of one set of weights to re-weight the posteriors. 
        With these weights we are able to pass the inference test. 
    }
    \label{fig:inf_lines}
\end{figure*}

When considering multiple chosen probabilities, $P_{\text{true}}$, and resulting coverage probabilities, $P_{\text{inf}}$, the results can be plot against each other.
The results of our inference test at $z = 5.4$ from 500 posteriors with true parameters randomly drawn from our priors are shown in the left panel of Figure \ref{fig:inf_lines}.
The grey shaded regions around our resulting line show the Poisson errors for our results.
Again we expect $P_{\text{true}} = P_{\text{inf}}$ which would give the red dashed line in this figure. 
To interpret this plot, first consider one point, for example $P_{\text{true}} \approx 0.6$. 
This represents the 60th percentile contour, which was calculated by the 60th percentile of the probabilities from the draws of the MCMC chain for each mock data set. 
Here, the true parameters fall within the 60th percentile contour only $\approx 50\%$ of the time. 
This implies that our posteriors are too narrow and should be wider such that the true model parameters will fall in the 60th percentile contour more often, so we are in fact underestimating our errors. 
We run this inference test at all $z$ considered in this work and found the deviation from the 1-1 line is worse at higher redshifts. 
See Appendix \ref{appendix:inference z6} for a discussion of the inference test at $z = 6$. 
We additionally run the inference test for mock data generated from a multi-variate Gaussian distribution in Appendix \ref{appendix: gauss data inf}. 

There has also been past work trying to correct posteriors that do not pass this coverage probability test \citep{prangle_2014, grunwald_2017, sellentin_2019}.
In this work, we are using the method of Hennawi et al. in prep. where we can calculate one set of weights for the MCMC draws that broaden the posteriors in a mathematically rigorous way.

A brief description of the reweighting method from Hennawi et al. in prep. is as follows.
Consider one data set which gives a corresponding posterior PDF. Initially we have:
\begin{equation}
\int d\hat{x} p(\hat{x}) H(p(\hat{x}) > p_0) = \alpha_0
\end{equation}
where $p(\hat{x})$ is the PDF of the posterior of some parameters $\hat{x}$,
$p_0$ is a chosen posterior probability, 
$H$ is the Heaviside function -- causing the integrand to be 0 when the probability is less than the given contour $p_0$, and $\alpha_0$ is the corresponding volume of the PDF where the probability of $\hat{x}$ is greater than $p_0$. 
This means that $\alpha_0 = 1 - C(p_0)$ where $C$ is the cumulative distribution function.

If we instead consider our MCMC chain used to estimate the posterior with $N_\text{samples}$ points each with probability $1 / N_{\text{samples}}$ in the chain we get:
\begin{equation}
\frac{1}{N_{\text{samples}}}\sum_i^{N_\text{samples}} H(p_i > p_0) = \frac{\text{\# of samples with }p > p_0}{N_{\text{samples}}} =  \alpha_0
\end{equation}
where the last equality comes from the fact that this sum is a Monte Carlo integral.

Consider the corresponding percentile, $P_{\text{true}}$, of this probability contour.
By definition, $C(P_{\text{true}}) = 1 - P_{\text{true}}$ (because the greatest values of the probability correspond to the smallest percentile contours). 
Thus we have:
\begin{equation}
\frac{1}{N_{\text{samples}}}\sum_i^{N_\text{samples}} H(p_i > p_0) =  P_{\text{true}}
\label{eq: monte_sum}
\end{equation}

However, as discussed above, after running an inference test what was thought of as the $P_{\text{true}}$ percentile contour is in reality the inferred percentile, $P_{\text{inf}}$, contour. Previous works \citet{sellentin_2019} suggested relabeling the $P_{\text{true}}$ contour as the $P_{\text{inf}}$ contour. However, another method to broaden (or condense) the probability contour is by using a set of weights. Consider re-writing equation \eqref{eq: monte_sum}  using weights, $w$:
\begin{equation}
\frac{1}{N_{\text{samples}}}\sum_i^{N_\text{samples}} w(x_i) H(p_i > p_0) \approx  f(P_{\text{true}})
\end{equation}
You can then consider multiple values of $P_{\text{true}}$ and absorb the factor of $\frac{1}{N_{\text{samples}}}$ into the weights:
\begin{equation}
A \boldsymbol{w} = f(\boldsymbol{P_{\text{true}}}) 
\end{equation}
where $A$ is matrix of only 1s and 0s from the Heaviside function, $\boldsymbol{w}$ is the vector of weights, and  $\boldsymbol{P_{\text{true}}}$ is the vector of probability contours considered. In fact, we can order the samples from the smallest probability value to the largest probability value such that $A$ is an upper triangular matrix. 
To guarantee the new weighted probability contours behave as they should statistically (i.e. the true value falls in the $P$-th percentile contour $P$\% of the time), we set $f(P_{\text{true}}) = P_{\text{inf}}$. This works because $P_{\text{inf}}$ is the measured statistically correct percentile contour for this $P_{\text{true}}$ value from the previous inference test. 
Therefore, we can compute weights by:
\begin{equation}
\boldsymbol{w} = A^{-1} \boldsymbol{P_{\text{inf}}}. 
\end{equation}

Note that this equation implies that we must run the inference test for the number of probability contours equal to the number of MCMC probability samples we have for each posterior. However, in practice we compute much fewer $P_{\text{inf}}$ values during the inference test and then interpolate this vector onto one with the number of MCMC samples we have.

Thus we are able to calculate one set of $(5000-500) \times 16 = 72000$ weights that would be used for all 500 mock data sets to broaden the posteriors and pass this inference test.
The weights calculated by this method, for a given set of MCMC chains, are unique. 
The line resulting from the inference test after calculating and using a set of weights is shown in the right panel of Figure \ref{fig:inf_lines}. 
This line clearly falls along the 1-1 line as expected so our calculated weights allow us to re-weight our posteriors into a statistically correct form. 
See Appendix \ref{appendix:inference z6} for a discussion of the re-weighting at $z = 6$. 

We show the re-weighted posteriors on $\lammfp$ and $\langle F \rangle$ in the bottom right part of Figure \ref{fig:corner_fit}.
The weights give greater importance to larger values of $\lammfp$ in the tail of values to the right, effectively broadening the posteriors and increasing the errors on the fit.
For the mock data set in Figure \ref{fig:corner_fit} the re-weighted marginalized posterior for $\lammfp$ gives $\lammfp = 30^{+14}_{-7}$ cMpc whereas before the inferred value was $30^{+10}_{-6}$ cMpc, so the new errors are $\sim 30\%$ larger. 
When looking at the 2D distribution in this panel we see that the weights do introduce an additional source of noise to the posterior distribution. 

This whole inference procedure is not the optimal and will not give the best constraints on $\lammfp$ possible from this statistic. 
The need to use re-weighting, or some method to correct our posteriors to pass an inference test, comes from our incorrect (though frequently used) assumption of a multivariate Gaussian likelihood. 
The values of the auto-correlation function at these high $z$ do not sufficiently follow a multivariate Gaussian distribution to justify this assumption, which should be a warning for other studies of the \lya forest at these $z$. 
Using a more correct form of the likelihood (such as a skewed distribution) or likelihood-free inference (such as approximate Bayesian computation as used in \citet{davies_2018_abc} or other machine learning methods) would lead to more optimal posteriors that better reflect the information in the auto-correlation function.
Therefore, future work on this inference procedure will improve the constraints on $\lammfp$.

\section{Results} \label{section: results}

\begin{table}
    \begin{tabular}{|c|c|cc|}
        \hline
        \multirow{2}{*}{$z$} & \multirow{2}{*}{Model $\lammfp$ (cMpc)} & \multicolumn{2}{c|}{Measured $\lammfp$ (cMpc)}        \\
                             &                                         & $R = 8800$                           & $R = 30000$    \\ \hline
        5.4                  & 39                                      & \multicolumn{1}{c|}{$40^{+27}_{-9}$} & $32^{+7}_{-5}$ \\
        5.5                  & 32                                      & \multicolumn{1}{c|}{$35^{+12}_{-6}$} & $33^{+6}_{-4}$ \\
        5.6                  & 26                                      & \multicolumn{1}{c|}{$28^{+7}_{-4}$}  & $27^{+5}_{-3}$ \\
        5.7                  & 20                                      & \multicolumn{1}{c|}{$22^{+7}_{-4}$}  & $20^{+4}_{-3}$ \\
        5.8                  & 16                                      & \multicolumn{1}{c|}{$18^{+6}_{-4}$}  & $16^{+3}_{-3}$ \\
        5.9                  & 12                                      & \multicolumn{1}{c|}{$14^{+5}_{-4}$}  & $13^{+3}_{-3}$ \\
        6.0                  & 9                                       & \multicolumn{1}{c|}{$12^{+6}_{-3}$}  & $11^{+4}_{-2}$ \\ \hline
    \end{tabular}
    \centering
    \caption{
    This table contains the results of analyzing the $\lammfp$ posteriors for the model value of the auto-correlation function.
    The mock data at each $z$ has the same value of $\lammfp$ as recorded in Table \ref{tab:central vals}. 
    The first column contains the modeled value of $\lammfp$ at each $z$ that was used in this measurement. 
    The next column contains the resulting measurements at each $z$ for $R = 8800$ data while the last column has the resulting measurements for $R = 30000$ data. 
    In general the trend of the errors is to initially decrease with increasing redshift and then stay about flat beyond $z = 5.7$. 
    This trend follows from the evolution in the true value of $\lammfp$ and the data set size at each $z$. 
    }
    \label{tab:measurements}
\end{table}

In order to consider the range of observational constraints possible from one set of $\lammfp$ and $\langle F \rangle$ values because of cosmic variance, we study the distribution of measurements for 100 mock data sets. 
For each $z$ we use the $\lammfp$ and $\langle F \rangle$ values reported in Table \ref{tab:central vals}.
Each mock data set is chosen by randomly selecting the appropriate number of skewers given the data set size at each redshift, and averaging the auto-correlation function for each individual skewer. 
For each mock data set, we perform MCMC as described in Section \ref{section: mcmc} and then re-weight the resulting posteriors following Section \ref{section: inf re-weight}.
Once we have the weights and the chains resulting from our inference procedure we can calculate the marginalized posterior for $\lammfp$. 

\begin{figure}
	\includegraphics[width=\columnwidth]{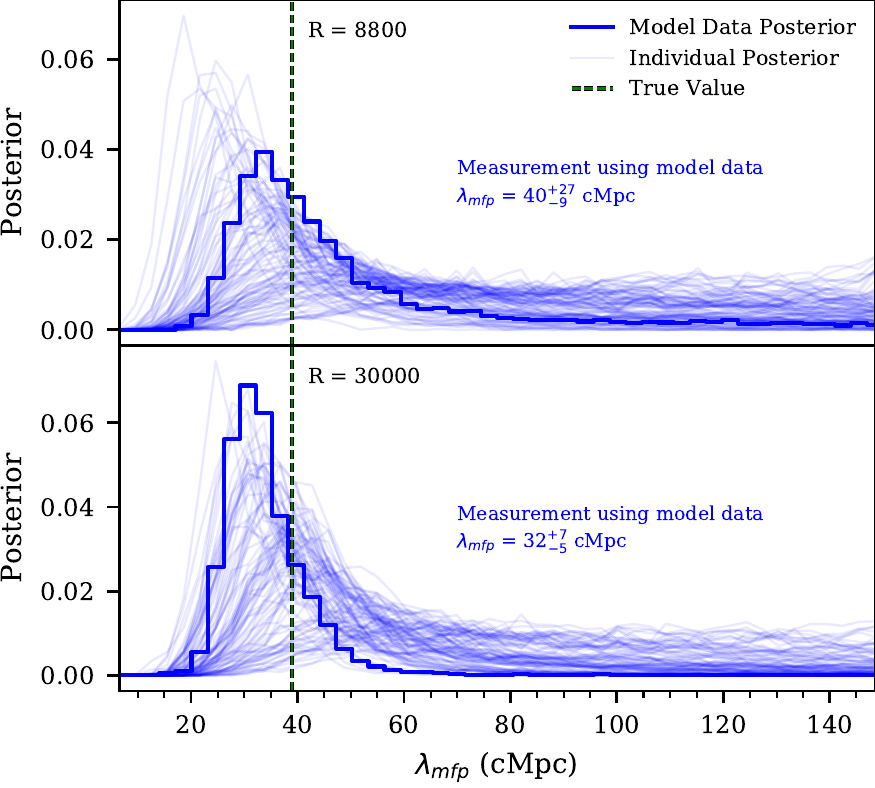}
    \caption{
        This figure shows 100 re-weighted posteriors of $\lammfp$ at $z = 5.4$ with true $\lammfp = 39$ cMpc and $\langle F \rangle = 0.0801$ (blue faint lines).
        The top panel shows the low-resolution $R = 8800$ results and the bottom panel shows the high-resolution $R = 30000$ results. 
        It also displays the re-weighted posterior (thick blue histogram) from the model's value of the auto-correlaiton function with the measurement of this average posteriors written in blue text.
        This demonstrates the different possible behaviors the posterior can have from our method. 
        Overall, the higher resolution data does produce more precise posteriors, including the average posterior which is seen in the higher peak and smaller reported errors. 
    }
    \label{fig:many_post}
\end{figure}

We calculate the marginalized re-weighted posteriors for 100 mock data sets at each $z$ and $R$. 
All 100 marginalized re-weighted posteriors are shown as the faint blue lines in Figure \ref{fig:many_post} at $z = 5.4$ for $R = 8800$ (top panel) and $R = 30000$ (bottom panel).
In addition to the randomly selected mock data sets, we computed the re-weighted posterior using the model value of the auto-correlation. 
This is shown as the blue histogram in Figure \ref{fig:many_post}. 
Using the model value as mock data is the ideal case and removes the luck of the draw from affecting the precision of this posterior. 
The measurement resulting from the model data is written in the blue text of this figure and the values at each $z$ and $R$ are reported in Table \ref{tab:measurements}. 

Figure \ref{fig:many_post} shows the results from all 100 mock data sets (blue lines) at $z = 5.4$ in order to visualize the various shapes of the resulting re-weighted posteriors.
These data sets all have the true values of $\lammfp = 39$ cMpc, $\langle F \rangle = 0.0801$, and a 64 quasar data set. 
The top panel shows the low-resolution $R = 8800$ results and the bottom panel shows the high-resolution $R = 30000$ results. 
The re-weighted histograms in Figure \ref{fig:many_post} are noisy, much like is seen in the bottom right panel of Figure \ref{fig:corner_fit}. 
This is a direct consequence of our re-weighting procedure and will be improved with further work on likelihood-free inference. 
There are roughly two behaviors of posteriors shown here: those that have a large peak at low values and those that are lower limits starting at low value and staying non-zero at our upper boundary of 150 cMpc. 
For both the lower resolution and higher resolution data, the model values of the auto-correlation function give posteriors with typical widths when compared to the mock data. 
Both model posteriors also contain the true value of $\lammfp$ within their $1 \sigma$ error bars.
Overall, the higher resolution data does produce tighter, more precise posteriors for both the model value and the mock data. 

Table $\ref{tab:measurements}$ reports the measurements that result from using the model values of the auto-correlation function as our data. 
This is an ideal scenario that removes luck of the draw from the resulting measurement. 
The first column contains the modeled value of $\lammfp$ at each $z$ that was used in this measurement, which also appear in Table \ref{tab:central vals}. 
The next column has the resulting measurements for $R = 8800$ data while the last column has the resulting measurement for the $R = 30000$ data. 
The errors initially decrease with increasing redshift and then stay about the same beyond $z = 5.7$. 
There are two important factors to consider when looking at this trend. 
First is the trend of the true value of $\lammfp$ with $z$ where
$\lammfp$ decreases with increasing $z$. 
The auto-correlation function is more sensitive to smaller values of $\lammfp$ as discussed in Section \ref{section: autocorr}. 
Briefly this is due to the fact that smaller $\lammfp$ produce greater fluctuations resulting in a larger signal. 
This means we would expect the results to get more precise and thus have smaller errors at higher $z$. 
The other factor is the size of the data set, which is greatest at the lowest $z$. 
We would expect the measurements to be less precise and thus have larger errors for the smaller data sets at high $z$. 
These effects combine resulting in the trend we see. 
When comparing the $R=8800$ and $R = 30000$ measurements, we find that the $R = 30000$ values are on average $40\%$ more precise. 
Note that it also appears that the measured values of $\lammfp$ are always biased high. 
However, these posterior distributions have tails to greater values of $\lammfp$ which causes the reported median measured value of $\lammfp$ to be greater than the most likely value of $\lammfp$.

\begin{figure*}
	\includegraphics[width=2\columnwidth]{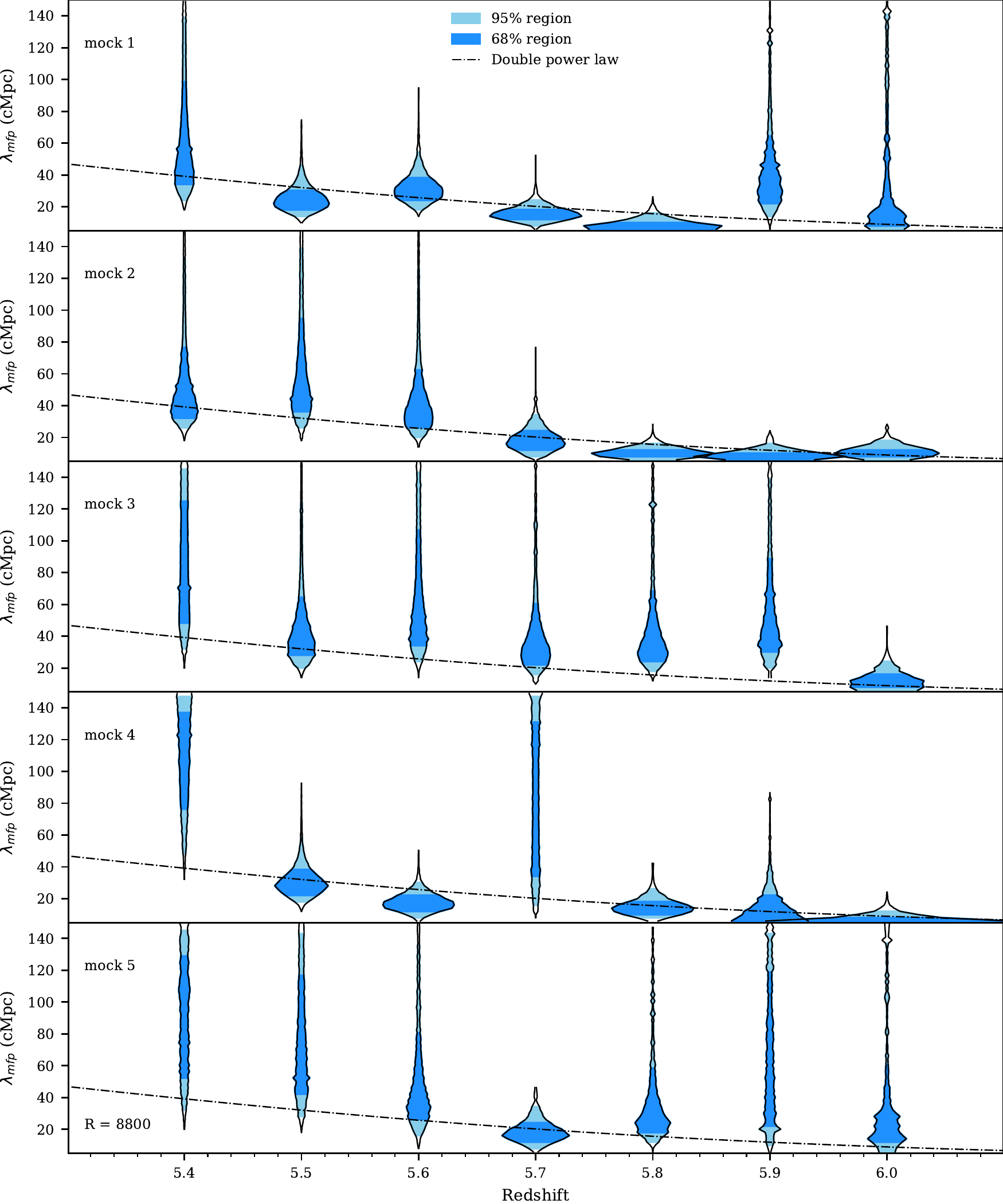}
    \caption{
        Each panel of this figure shows one posterior for a different randomly selected low-resolution ($R = 8800$) mock data set at each $z$. 
        For each posterior, the light blue shaded region demarcates the 2.5th and 97.5th percentile of the MCMC draws while the darker blue shaded region demarcates the 17th and 83rd percentile of the MCMC draws. 
        The black dot dashed line shows the double power law from equation \eqref{eq:double power law} and Figure \ref{fig:mfp_model_evolution}. 
        The behavior of each posterior at the different $z$ is determined by the luck of the draw when selecting the mock data, the true $\lammfp$ value at each $z$, and the data set size at each $z$. 
        The true $\lammfp$ values and data set sizes are reported in Table \ref{tab:central vals}.
    }
    \label{fig:violin_plot}
\end{figure*}

In order to visualize the differences between measurements at different redshifts, we plot the results for five random mock data sets with $R = 8800$ in Figure \ref{fig:violin_plot}. 
Each violin is the re-weighted marginalized posterior for one randomly selected mock data set at the corresponding redshift. 
The light blue shaded region demarcates the 2.5th and 97.5th percentiles (2$\sigma$) of the MCMC draws while the darker blue shaded region demarcates the 16th and 84th percentiles (1$\sigma$) of the MCMC draws. 
The dot dashed line is the double power law, equation \eqref{eq:double power law}, which we used to determine the true $\lammfp$ evolution as shown in Figure \ref{fig:mfp_model_evolution}. 

Looking at the posteriors for a given redshift (one column in the figure), the only difference between the posteriors is the random mock data set drawn. 
This still produces different precision results as seen in Figure \ref{fig:many_post} for $z = 5.4$. 
There are then three differences between mock data sets shown for a given panel. 
First is again the mock data is chosen at random so there will be fluctuations in the precision with the luck of the draw. 
The mock data at different redshift also have different true $\lammfp$ values, shown in the dot-dashed black line, where the smallest $\lammfp$ value is at the highest $z$. 
The auto-correlation function is most precise at small inferred $\lammfp$ values which are more likely at the highest $z$.
Additionally, the redshifts each have different data set sizes, as reported in Table \ref{tab:central vals}. 
The highest redshifts have the smallest data set sizes, leading to greater scatter in the precision of the posteriors. 
Again, the individual posteriors are noisy, resulting from the re-weighting procedure as described in Section \ref{section: inf re-weight}.  

Here all the mock data sets are at our lower-resolution, $R = 8800$, which was chosen to mimic the existing XQR-30 data. 
In Appendix \ref{appendix: high-z results} we discuss the same plot (Figure \ref{fig:violin_plot_appendix}) but with the higher resolution, $R = 30000$, data. 
The only difference between the data used in Figure \ref{fig:violin_plot} and Figure \ref{fig:violin_plot_appendix} is the resolution of the mock data. 
The randomly chosen mock data sets, the data set sizes, and the true values are the same.

\section{Conclusions} \label{section: conclusion}

In this work we have investigated to what precision $\lammfp$ can be constrained using the auto-correlation function of \lya forest flux in quasar sightlines. 
Overall, we found that the auto-correlation function is sensitive to the value of $\lammfp$ across multiple redshift bins and for realistic mock data with both high and low resolution. 
We computed the marginalized re-weighted posterior for $\lammfp$ for 100 mock data sets with properties similar to the XQR-30 extended data set at $5.4 \leq z \leq 6.0$
We additionally considered 100 mock data sets with $R = 30000$, over three times greater than XQR-30 data resolution. 
The re-weighted posterior showed a variety of behaviors based on the luck of the draw of the mock data chosen, the true value of $\lammfp$ for the mock data, and the data set size at each $z$. 

We considered an ideal data set which had the model value of the auto-correlation function, effectively removing the luck of the draw from our measurement.
The error on these measurements for both the high resolution and low resolution data initially got smaller (more precise) with increasing redshift then stayed about the same beyond $z = 5.7$. 
This followed from the changing true value of $\lammfp$ and the size of the data set at each $z$.
Small values of $\lammfp$ lead to greater fluctuations in the UVB and thus produce an increased signal in the auto-correlation function. 
This makes the auto-correlation function more sensitive to smaller values of $\lammfp$ than larger values of $\lammfp$ where the fluctuations are smaller. 
This work has opened up the possibility for future measurements of $\lammfp$ with the auto-correlation function by quantifying the sensitivity of this method. 

Of particular interest is the measurement at $z = 6.0$, where recent measurements imply a rapid evolution of $\lammfp$. 
For our ideal model data at $z = 6.0$ with $R = 8800$, we get $\lammfp = 12^{+6}_{-3}$ cMpc where the true value we modeled was $\lammfp = 9$ cMpc. 
In comparison, the measurement from \citet{becker_2021} at $z = 6.0$ is $0.75^{+0.65}_{-0.45}$ proper Mpc (or $5.25^{+4.25}_{-3.15}$ cMpc).
Thus our ideal measurement with this new statistical method has comparable error bars as those from \citet{becker_2021}. 
We therefore expect that a measurement using this technique on real data will provide a competitive, secondary check on the value of $\lammfp$ at $z = 6.0$. 
Additionally, we have shown that our method can be applied to multiple fine redshift bins from $5.4 \leq z \leq 6.0$ to precisely constrain the evolution of $\lammfp$. 

Note that our procedure uses a multi-variate Gaussian likelihood, MCMC, and a set of weights for the MCMC chains that ensures our posteriors pass an inference test.
The original failure of our procedure to pass an inference test is likely due to the incorrect assumption that the auto-correlation function follows a multi-variate Gaussian distribution, as discussed in Appendix \ref{appendix:multi gaussian}. 
This result should caution against using a multi-variate Gaussian likelihood with other statistics, such as the power spectrum, when making measurements at $z > 5$ as the same issue of non-Gaussian data likely applies. 
This is especially concerning if the low value of $\lammfp$ with high corresponding fluctuations in the UVB at high-$z$ holds true. 
In the future, better likelihoods or likelihood-free inference will allow for a more optimal inference procedure (see e.g. \citet{davies_2018_abc} or \citet{alsing_2019}). 
This will lead to tighter constraints on $\lammfp$ from the auto-correlation function. 

For this work, we used the method of \citet{davies_furlanetto_2016} to generate the UVB boxes as described in section \ref{section: sim data uvb model}. 
This assumes a fixed source model which could potentially prove to be incorrect. 
For example, if fainter galaxies had higher escape fractions it would reduce the strength of UVB fluctuations at fixed $\lammfp$, also reducing the auto-correlation signal. 
This would bias $\lammfp$ measurements through this method from real data compared to these models (though it is consistent for our mock data generated from our models). 
We leave a detailed consideration of the effect of other source models to future work.

Our work also discussed the effect of the current limitations in modeling the UVB and Ly$\alpha$ forest on the auto-correlation function. 
Namely, our UVB boxes are not correlated with the density of our \texttt{Nyx} simulation box, where in reality these quantities are physically correlated. 
We considered the effect of a correlated UVB in Section \ref{section:mfp_density_match}. 
We found that the correlation between high density areas with increased UVB values would reduce the auto-correlation signal for a fixed $\lammfp$ on small scales, since higher density leads to reduced transmission. 
This would again bias a measurement from real data, where these correlations would exist, because the true signal for a given $\lammfp$ should be lower than it is in our models, which mimics a model with a larger $\lammfp$ value. 
However, this comparison was done in a small box (40 cMpc h$^{-1}$) which suppresses UVB fluctuations on all scales as is also discussed in Section \ref{section:mfp_density_match}.
Suppressing fluctuations in the UVB causes the auto-correlation signal to be lower in these boxes.
Thus in this comparison the signal is smaller from the density correlations but the UVB fluctuations are also under-estimated due to the box size. 
The existence of both of these effects means that we were not able to quantify any potential bias from the uncorrelated UVB boxes. 
The mock data used in this work is generated in the same ways as the models they are compared to, so the measurements here are self-consistent.
However, any attempts to compare these models with actual data will need to take into account the effect of using an uncorrelated UVB in the modeling. 
Thus, future work on UVB models will be necessary before observational $\lammfp$ constraints can be produced.

Another potential physical impact on the auto-correlation signal is fluctuation in the temperature of the IGM. 
\citet{onorbe_2019} showed that fluctuations in the temperature of the IGM impacted the largest scales of the power spectrum at $z > 5$. 
We therefore would conclude these fluctuations would also impact the auto-correlation function, which is the Fourier transform of the power spectrum. 
However \citet{onorbe_2019} also considered a fluctuating UVB and found that this effectively cancelled out the impact of the thermal fluctuations on the largest scales of the power spectrum.
We leave further work on the impact of temperature fluctuations along with UVB fluctuations to future work. 

Continuum errors will effect the measurement of the auto-correlation on larger scales which are less important than the small scales when considering $\lammfp$. 
The reconstruction done in \citet{bosman_2021_data} is shown to reconstruct the continuum within $8\%$. 
Additionally, \citet{eilers_2017} showed that continuum errors had minimal effect on the shape of the normalized flux PDF at $z = 5$ where transmission is low. 
We have left a detailed exploration of the effect of continuum errors on the auto-correlation function for future work. 

We also note that there is additional $z > 5$ \lya forest data in telescope archives with lower SNR that could be used in our analysis. 
Here we limited the consideration to mock XQR-30 data (and a high-resolution analog) but will consider the impact of adding noisier data in future work. 

The value of $\lammfp$ and its evolution at high $z$ is important for understanding reionization. 
Measuring $\lammfp$ at high $z$ is a difficult task that so far has been restricted to two redshift bins at $z > 5$.
This work has shown that the auto-correlation function of the \lya forest flux provides a new, competitive way to constrain $\lammfp$ in multiple redshift bins at $z \geq 5.4$.

\section*{Acknowledgements}


We acknowledge helpful conversations with the ENIGMA group at UC Santa Barbara and Leiden University. 
This project has received funding from the European Research Council (ERC) under the European Union’s Horizon 2020 research and innovation program (grant agreement No 885301). 
JFH acknowledges support from the National Science Foundation under Grant No. 1816006 .
This research also used resources of the National Energy Research Scientific Computing Center (NERSC), a U.S. Department of Energy Office of Science User Facility located at Lawrence Berkeley National Laboratory, operated under Contract No. DE-AC02-05CH11231.

\section*{Data Availability}

The simulation data analyzed in this article will be shared on reasonable request to the corresponding author.



\bibliographystyle{mnras}
\bibliography{corr_forecast_paper} 



\appendix

\section{Convergence of the Covariance Matrices} \label{appendix:converge}

\begin{figure}
	\includegraphics[width=\columnwidth]{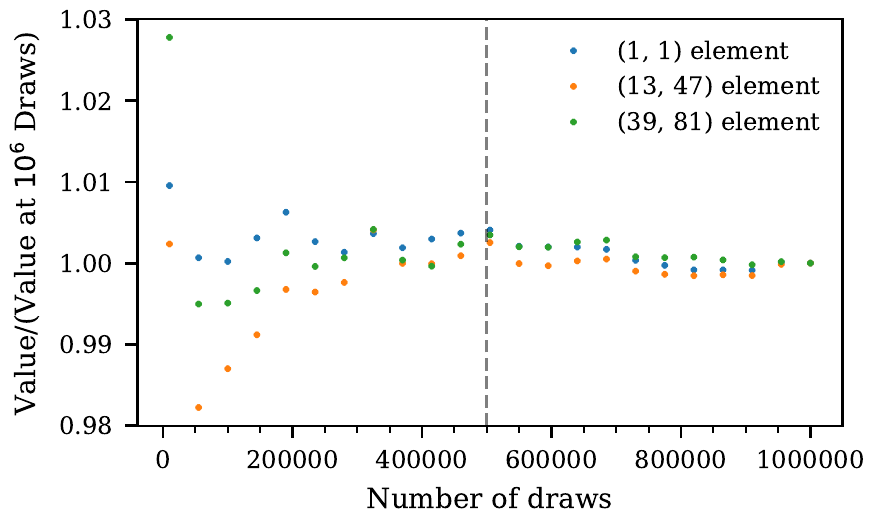}
    \caption{
        This figure shows the behavior of three elements of the model covariance matrix ($z = 6$, $R = 8800$, $\lammfp = 9$ cMpc, and $\langle F \rangle = 0.0089$) for different numbers of mock draws. 
        At all values of the number of mocks considered, the covariance elements fall within 3\% of their final value. 
        By around $\sim 200000$ draws, all of the values fall within 1\% of the final value. 
        For this reason, we believe using 500000 mock draws is sufficient to generate the covariance matrices used in this study. 
        500000 mock draws is represented by the vertical dashed black line. 
    }
    \label{fig:covar_conv}
\end{figure}

We calculate the covariance matrices for our models with mock draws, as defined in equation \eqref{eq:covariance}.
Using mock draws is inherently noisy and it should converge as $1 / \sqrt{N}$ where $N$ is the number of draws used. 
As stated in the text, we used 500000 mock draws. 
To check that this number is sufficient to minimize the error in our calculation, we looked at the behavior of elements of one covariance matrix in Figure \ref{fig:covar_conv}. 
This covariance matrix is for the model with $z = 6$, $R = 8800$, $\lammfp = 9$ cMpc, and $\langle F \rangle = 0.0089$, chosen because $z = 6$ has the lowest ``true" $\lammfp$ value which would lead to the largest fluctuations in the UVB. 
The values in the plot have been normalized to 1 at $10^6$ draws. 
The three elements have been chosen such that there is one diagonal value and two off-diagonal values in different regions of the matrix. 
Looking at the correlation matrix in Figure \ref{fig:covar_correlation} (which is for a different model but the qualitative behavior is the same for this model) we see that at all values both on and off the diagonal of the correlation matrix are high and positive, so we expect the convergence for all elements to be roughly the same. 
At all values of the number of mock draws considered, the covariance elements fall within 3\% of their final value. 
By around $\sim 200000$ draws, the values fall within 1\% of the final value. 
For this reason, we believe using 500000 mock draws is sufficient to generate the covariance matrices used in this study. 
In Figure \ref{fig:covar_conv}, 500000 mock draws is represented by the vertical dashed black line.

\section{Non-Gaussian distribution of the values of the auto-correlation function} \label{appendix:multi gaussian}

For our inference, we used the multi-variate Gaussian likelihood defined in equation \eqref{eq:gauss_like}. 
This functional form assumes that the distribution of mock draws of the auto-correlation function is Gaussian distributed about the mean for each bin.
In order to visually check this we will look at the distribution of mock draws from two bins of the auto-correlation function for two different models. 

\begin{figure}
	\includegraphics[width=\columnwidth]{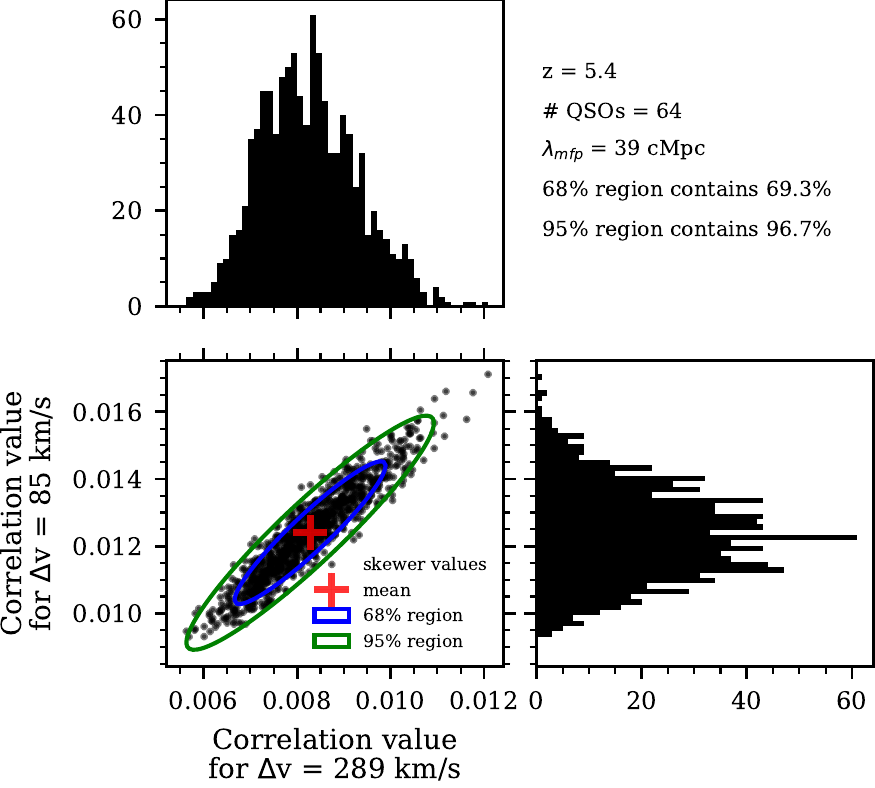}
    \caption{
        This figure shows the distribution 1000 mock draws from two bins of the auto-correlation function ($\Delta v = \SI{85.0}{\kilo\meter\per\second}$ and $\Delta v = \SI{289.0}{\kilo\meter\per\second}$) for one model ($z = 5.4$, $R = 8800$, $\lammfp = 39$ cMpc, and $\langle F \rangle = 0.0801$).
        The top panel shows the distribution of only the $\Delta v = \SI{289.0}{\kilo\meter\per\second}$ bin while the right panel shows the distribution of only the $\Delta v = \SI{85.0}{\kilo\meter\per\second}$ bin. 
        The blue (green) circle represents the 68\% (95\%) ellipse calculated from the covariance matrix calculated for this model from equation \eqref{eq:covariance}. 
        The red plus shows the calculated mean. 
        Additionally the percent of mock draws that fall within each of these contours is written in the top right. 
        Both the 1D and 2D distributions seem relatively well described by a Gaussian distribution. 
        In the 2D plot, there are more points outside the 95\% contour to the top right than on any other side but it is not extreme. 
    }
    \label{fig:app_gaus_corner_z54}
\end{figure}

\begin{figure}
	\includegraphics[width=\columnwidth]{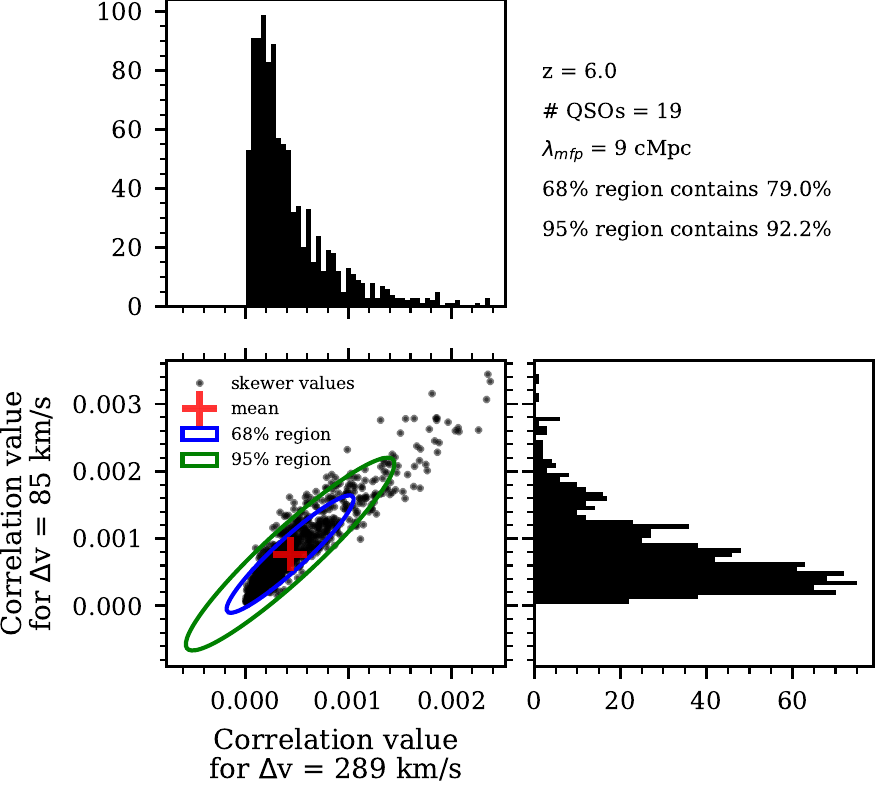}
    \caption{
        This figure shows the distribution 1000 mock draws from two bins of the auto-correlation function ($\Delta v = \SI{85.0}{\kilo\meter\per\second}$ and $\Delta v = \SI{289.0}{\kilo\meter\per\second}$) for one model ($z = 6$, $R = 8800$, $\lammfp = 9$ cMpc, and $\langle F \rangle = 0.0089$).
        The top panel shows the distribution of only the $\Delta v = \SI{289.0}{\kilo\meter\per\second}$ bin while the right panel shows the distribution of only the $\Delta v = \SI{85.0}{\kilo\meter\per\second}$ bin. 
        The blue (green) circle represents the 68\% (95\%) ellipse calculated from the covariance matrix calculated for this model from equation \eqref{eq:covariance}. 
        The red plus shows the calculated mean. 
        Additionally the percent of mock draws that fall within each of these contours is written in the top right. 
        Both the 1D and 2D distributions do not seem to be well described by a Gaussian with 79.0\% of the mock draws falling within the 68\% contour and 92.2\% of the mock draws falling within the 95\% contour.
    }
    \label{fig:app_gaus_corner_z6}
\end{figure}

Both Figures \ref{fig:app_gaus_corner_z54} and \ref{fig:app_gaus_corner_z6} show the distribution of 1000 mock data sets from the velocity bins of the auto-correlation function with $\Delta v = \SI{85.0}{\kilo\meter\per\second}$ and $\Delta v = \SI{289.0}{\kilo\meter\per\second}$.
The bottom left panels show the 2D distribution of the auto-correlation values from these bins. 
The blue (green) ellipses represents the theoretical 68\% (95\%) percentile contour calculated from the covariance matrix calculated for each model from equation \eqref{eq:covariance}. 
The red crosses shows the calculated mean. 
The top panels show the distribution of only the $v = \SI{289.0}{\kilo\meter\per\second}$ bins while the right panels show the distribution of only the $v = \SI{85.0}{\kilo\meter\per\second}$ bins. 

Figure \ref{fig:app_gaus_corner_z54} shows mock values of two bins of the auto-correlation function for the model at $z = 5.4$ with  $R = 8800$, $\lammfp = 39$ cMpc and $\langle F \rangle = 0.0801$. 
These mock data sets consist of 64 quasar sightlines of length $\Delta z = 0.1$.
Both the 1D and 2D distributions seem relatively well described by Gaussian distributions by eye though they do show some evidence of non-Gaussian tails to larger values. 
The number of points falling in each contour both fall within 2\% of the expected values. 
In the bottom left panel with the 2D distribution there are more mock values falling outside the 95\% contour to the top right (higher values) than in any other direction.
For this reason the distribution is not exactly Gaussian but a Gaussian visually appears as an acceptable approximation.

Figure \ref{fig:app_gaus_corner_z6} shows mock values of two bins of the auto-correlation function for the model at $z = 6$ with $R = 8800$, $\lammfp = 9$ cMpc, and $\langle F \rangle = 0.0089$. 
These mock data sets consist of 19 quasar sightlines of length $\Delta z = 0.1$.
In both the top and right panels, which show the distribution of values for one bin of the auto-correlation function, it is clear that the distribution of mock draws is skewed and a Gaussian is not a good approximation for the distributions. 
This is quantified by the percent of points in the two ellipses from the bottom left panel labeled in the top right with 79.0\% of the mock draws falling within the 68\% contour and 92.2\% of the mock draws falling within the 95\% contour.
The points outside of the contours are highly skewered towards the top right (higher values). 
It is only possible for the auto-correlation function to be negative due to noise, which generally averages to very small values approaching zero at the non-zero lags of the auto-correlation function. 
However real fluctuations in the UVB cause the positive fluctuations, making them much more likely and cause the resulting skewed distribution at high $z$ where the overall signal is closer to zero. 

Figures \ref{fig:app_gaus_corner_z54} and \ref{fig:app_gaus_corner_z6} show the changing distribution of the auto-correlation value with $\lammfp$, $\langle F \rangle$, and mock data set size. 
There is a greater deviation from a multi-variate Gaussian distribution at higher $z$.
It is possible that adding additional sightlines will cause the auto-correlation function to better follow a multi-variate Gaussian distribution due to the central limit theorem, though investigating this in detail is beyond the scope of the paper. 
The incorrect assumption of the multi-variate Gaussian likelihood thus contributes to the failure of our method to pass an inference test as discussed in Section \ref{section: inf re-weight} for $z = 5.4$ and Appendix \ref{appendix:inference z6} for $z = 6$. 
For our final constraints, we calculated weights for our MCMC chains such that the resulting posteriors do pass our inference test, as discussed in Section \ref{section: inf re-weight}.
The whole method of assuming a multi-variate Gaussian then re-weighting the posteriors in non-optimal and future work using a more correct likelihood or likelihood-free inference will improve our results.

\section{Inference test at high redshift} \label{appendix:inference z6}

\begin{figure*}
    \centering
    \begin{subfigure}[t]{0.49\textwidth}
        \centering
        \includegraphics[width=\linewidth]{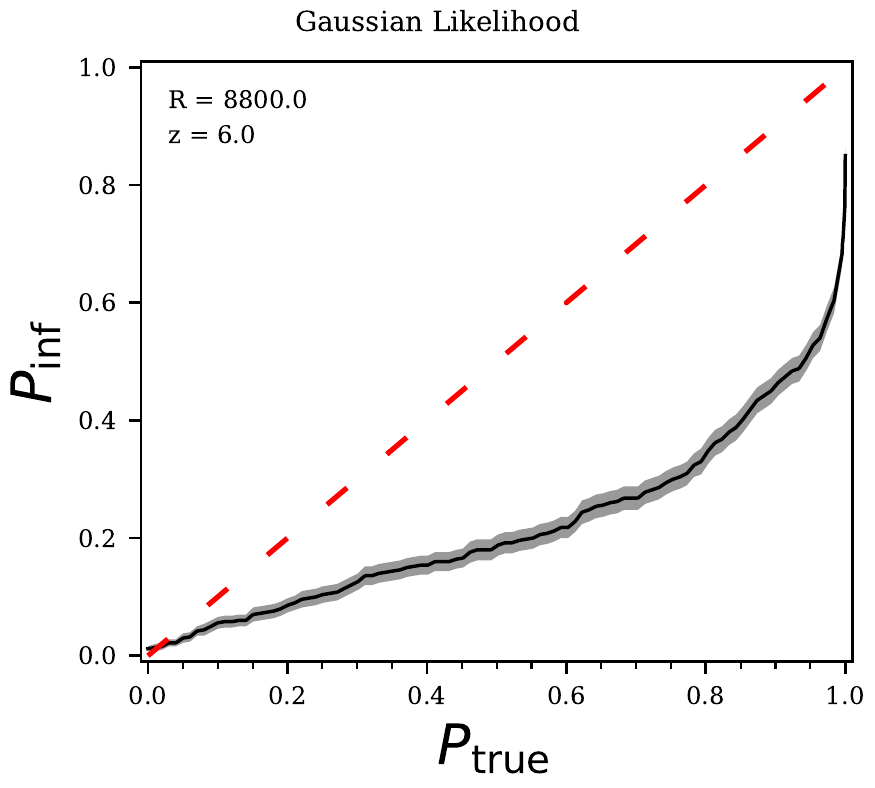} 
    \end{subfigure}
    \hfill
    \begin{subfigure}[t]{0.49\textwidth}
        \centering
        \includegraphics[width=\linewidth]{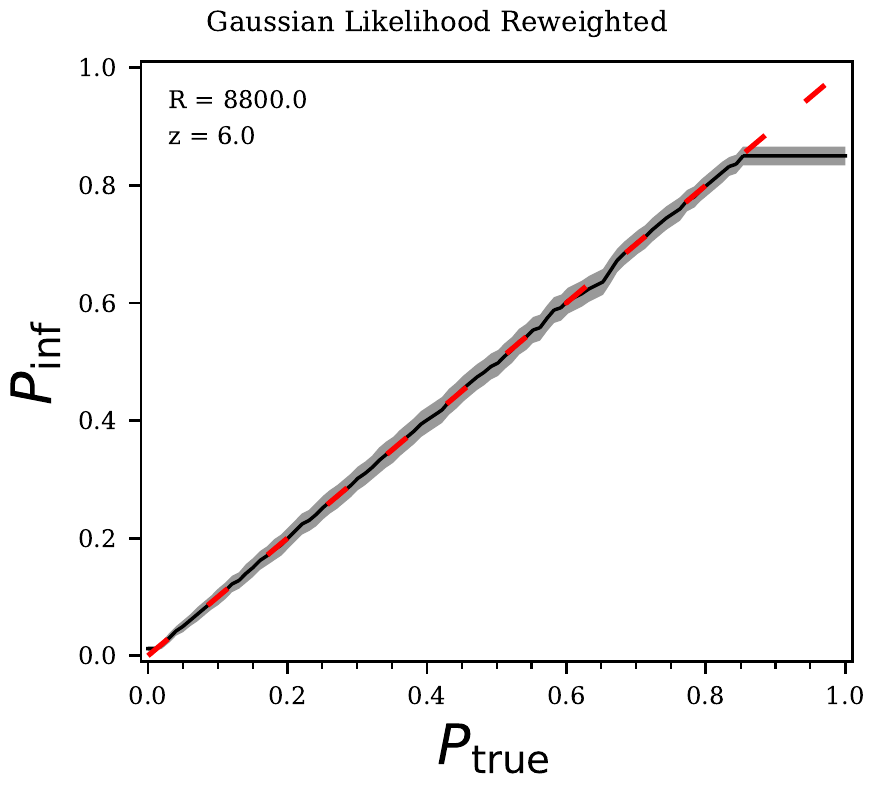} 
    \end{subfigure}
    \caption{
        The left panel of this figure shows the coverage plot resulting from the inference test from 500 models at $z = 6$ and $R = 8800$ drawn from our priors on $\lammfp$ and $\langle F \rangle$.
        Here we see that the true parameters for the models fall above the 60th percentile in the MCMC chain $\sim 20\%$ of the time, for example. 
        The right panel of this figure shows the coverage plot resulting from the inference test with the use of one set of weights to re-weight the posteriors. 
        With these weights the true parameters for the models fall on the $P_{\text{inf}} = P_{\text{true}}$ line up to $P_{\text{true}} \sim 0.8$. This is because the original coverage plot was only able to reach $P_{\text{inf}} \sim 0.8$ so our re-weighting could only match up to this value. 
    }
    \label{fig:inf_lines_z6}
\end{figure*}

Here we present the results of the inference test at $z = 6$. 
This calculation was done following the procedure described in Section \ref{section: inf re-weight}. 
Figure \ref{fig:inf_lines_z6} shows the results for $z = 6$ and can be compared to the $z = 5.4$ results in Figure \ref{fig:inf_lines}. 
The left panel here shows the initial coverage plot which deviates greatly from the expected $P_{\text{inf}} = P_{\text{true}}$ line, much more so than the $z = 5.4$. 
This likely comes from a greater deviation from the assumption of a multi-variate Gaussian likelihood as described in Appendix \ref{appendix:multi gaussian}. 
The $z = 6$ mock data show highly skewed distributions that are not well described by a Gaussian likelihood. 

This initial coverage plot only ever reaches a value of $P_{\text{inf}} \approx 0.8$, which becomes an issue for the re-weighting. 
In the right panel of Figure \ref{fig:inf_lines_z6} the re-weighted inference line thus still only able to reach $P_{\text{inf}} \approx 0.8$ creating a plateau in the line once it reaches this value. 
One way to reach higher values is to increase the number of steps in the MCMC chain. We tried to triple the number of steps but did not see much improvement in the inference test. 
For computational reasons we stick with the numbers used at other redshifts resulting in 72000 total steps as described in Section \ref{section: inf re-weight}.
This plateau at $P_{\text{true}} = 0.8$ means that our 1-$\sigma$ (68th percentile) contours are robust but our 2-$\sigma$ (95th percentile) contours are underestimated since we can only correct up to $\sim 80$th percentile. 

The inference lines at other redshifts are available upon request. 
For $5.4 \leq z \leq 5.8$ the coverage plots after re-weighting do not plateau, like the re-weighted coverage plot shown in Figure \ref{fig:inf_lines}. 
Both the $z = 6.0$ and the $z = 5.9$ coverage plots plateau after re-weighting, like that in Figure \ref{fig:inf_lines_z6}. 
This means our re-weighted posteriors at $z = 5.9$ and $z = 6$ may still need additional work to further enlarge probability contours above the value of the plateau.

\section{Gaussian data inference test} \label{appendix: gauss data inf}

\begin{figure*}
    \centering
    \begin{subfigure}[t]{0.49\textwidth}
        \centering
        \includegraphics[width=\linewidth]{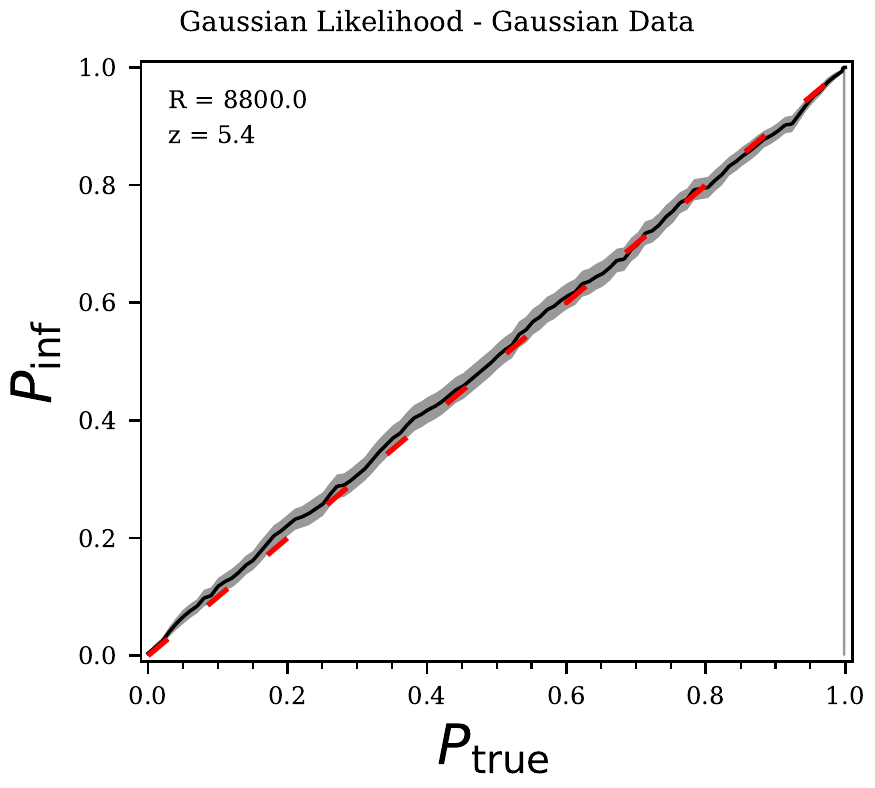} 
    \end{subfigure}
    \hfill
    \begin{subfigure}[t]{0.49\textwidth}
        \centering
        \includegraphics[width=\linewidth]{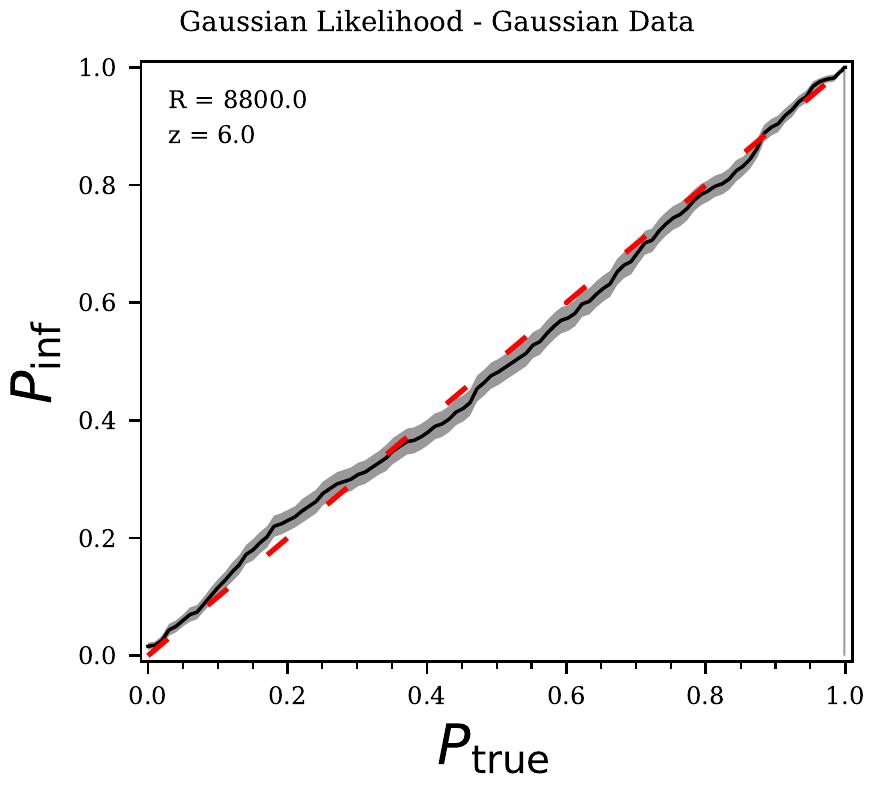} 
    \end{subfigure}
    \caption{
        Both panels of this figure shows the coverage plot resulting from the inference test from 500 data sets generated by randomly drawing points from the mean model and covariance matrix. 
        The the means and covariance matrices used come from $z = 5.4$ and $R = 8800$ in the left panel and $z = 6.0$ and $R = 8800$ in the right panel. 
        The true parameter values for both panels were drawn from our priors on $\lammfp$ and $\langle F \rangle$.
        In both panels, the Gaussian mock data produced inference lines that fall on top of the 1-1 line within errors, as expected for the statistically correct posteriors. 
    }
    \label{fig:inf_lines_gauss}
\end{figure*}

As shown in Appendix \ref{appendix:multi gaussian}, the distribution of mock values of the auto-correlation function is not exactly Gaussian distributed. 
In order to confirm the failure of our mock data to pass an inference test (as discussed in Section \ref{section: inf re-weight} and Appendix \ref{appendix:inference z6}) comes from the use of a multi-variate Gaussian likelihood, we generate Gaussian distributed data and run inference tests. 
For one value of $\lammfp$ and $\langle F \rangle$, we randomly generate a mock data set from a multi-variate Gaussian with the given mean model and covariance matrix that we calculated for our mock data in Section \ref{section: autocorr}. 
We can then continue with the inference test as described in Section \ref{section: inf re-weight}. 
The results for this inference test for $z = 5.4$ and $z = 6.0$ (both with $R = 8800$) are shown in Figure \ref{fig:inf_lines_gauss}. 
Here both redshifts inference lines fall along the 1-1 line that is expected for all probability contour, $P_{\text{true}}$, values. 
This behavior is also seen at the other redshifts and $R = 30000$. 
The fact that perfectly Gaussian data passes an inference test with the same likelihood, priors, and method as was used on mock data confirms that the failure of our mock data to pass an inference test is due to the non-Gaussian distribution of the mock data.

\section{High-resolution results} \label{appendix: high-z results}

\begin{figure*}
	\includegraphics[width=2\columnwidth]{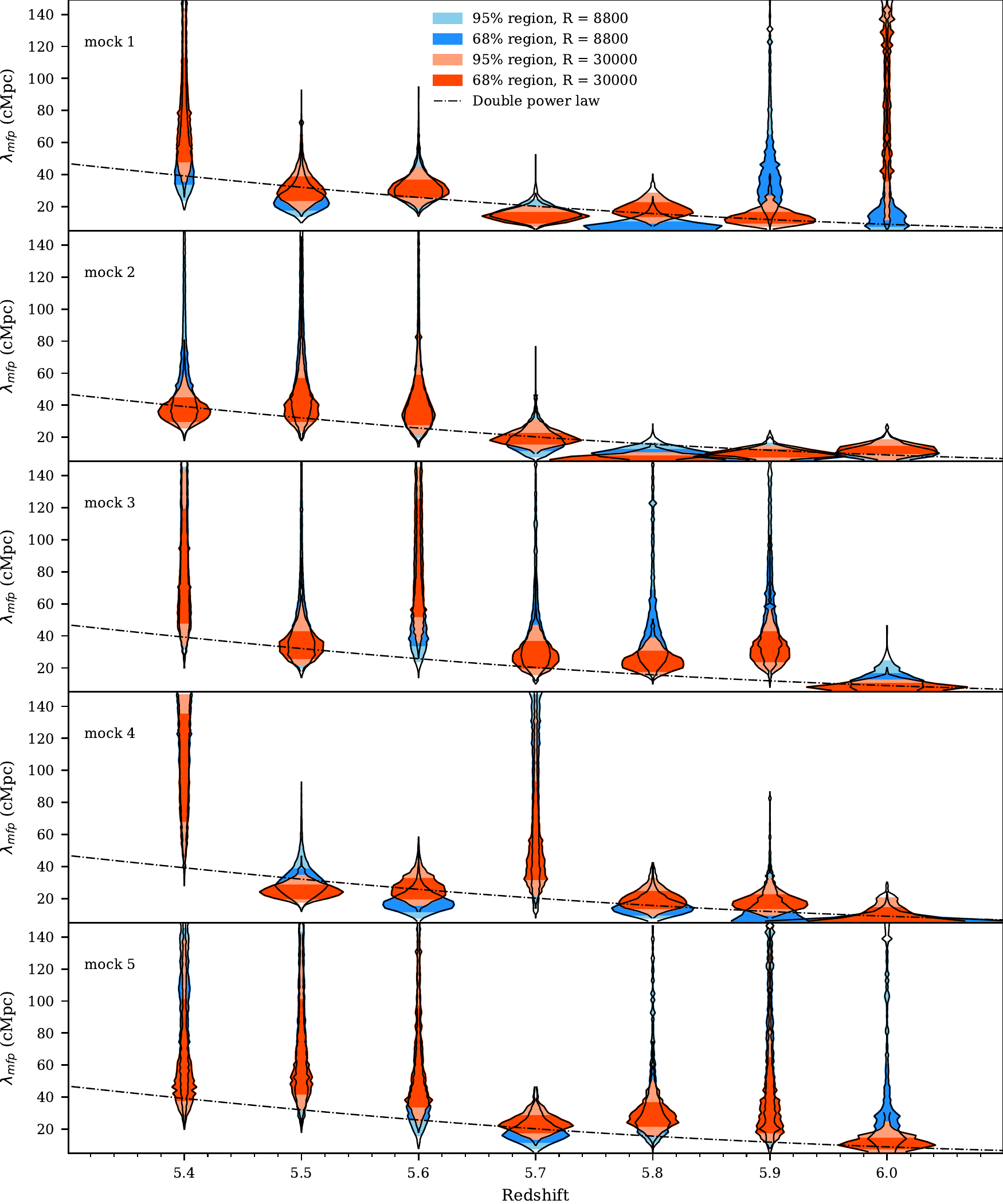}
    \caption{
        Each panel of this figure shows one posterior for a different randomly selected high-resolution ($R = 30000$) mock data set at each $z$ in shades of red. 
        Note that the low-resolution ($R = 8800$) mock data posteriors are plot below the high resolution posteriors in blue.
        For each posterior, the light red shaded region demarcates the 2.5th and 97.5th percentile of the MCMC draws while the darker red shaded region demarcates the 16th and 84th percentile of the MCMC draws. 
        The black dot dashed line shows the double power law from equation \eqref{eq:double power law} and Figure \ref{fig:mfp_model_evolution}. 
        The behavior of each posterior at the different $z$ is determined by the luck of the draw when selecting the mock data, the true $\lammfp$ value at each $z$, and the data set size at each $z$. 
        The true $\lammfp$ values and data set sizes are reported in Table \ref{tab:central vals}.
    }
    \label{fig:violin_plot_appendix}
\end{figure*}

In section \ref{section: results} we only show the posteriors for multiple mock data sets at different redshifts for $R = 8800$. 
Here we present the same results but for mock data with $R = 30000$. 
Each violin plot in Figure \ref{fig:violin_plot_appendix} is the re-weighted marginalized posterior for one randomly selected mock data set at the corresponding redshift. 
The light red shaded region demarcates the 2.5th and 97.5th percentiles (2$\sigma$) of the MCMC draws while the darker red shaded region demarcates the 16th and 84th percentiles (1$\sigma$) of the MCMC draws. 
Beneath the red violins are blue violins for the posteriors for the same data with $R = 8800$ as shown in  Figure \ref{fig:violin_plot}.
The dot dashed line is the double power law, equation \eqref{eq:double power law}, which we used to determine the true $\lammfp$ evolution as shown in Figure \ref{fig:mfp_model_evolution}. 
The random mock data selected for this figure matches exactly with the random mock data used to make Figure \ref{fig:violin_plot}. 
The only difference between the data used in these two figures is the resolution. 
Generally, the posteriors from the $R = 30000$ data shown in Figure \ref{fig:violin_plot_appendix} are more precise than those from the $R = 8800$ data.

Again, looking at the posteriors for a given redshift (one column in the figure), the only difference between the posteriors is the random mock data set drawn. 
These results still have varying precision as is expected from luck of the draw with the mock data sets. 
There are then three differences between mock data sets shown for a given panel. 
First is the same as the difference between mocks at one redshift: the mock data is chosen at random so there is just the luck of the draw. 
The mock data at each redshift also vary with the true $\lammfp$ value, shown in the dot-dashed black line, where the smallest $\lammfp$ value is at the highest $z$. 
The auto-correlation function is most precise at small inferred $\lammfp$ values which are more likely at the highest $z$.
Additionally, the redshifts each have different data set sizes, as reported in Table \ref{tab:central vals}. 
The highest redshifts have the smallest data set sizes, leading to greater scatter in the precision of the posteriors. 
Again, the individual posteriors are noisy, resulting from the re-weighting procedure as described in Section \ref{section: inf re-weight}.

\bsp	
\label{lastpage}
\end{document}